\documentclass[12pt]{iopart}

\usepackage{graphicx}
\usepackage{iopams}
\usepackage{bm}
\usepackage{cite}
\usepackage{dsfont}

\expandafter\let\csname equation*\endcsname\relax

\expandafter\let\csname endequation*\endcsname\relax

\usepackage{amsfonts}
\usepackage{amssymb}
\usepackage{amsmath}
\usepackage{bbold}
\usepackage{lipsum}

\usepackage[linktocpage=true]{hyperref}
\hypersetup{
    colorlinks,
    citecolor=blue,
    filecolor=blue,
    linkcolor=blue,
    urlcolor=blue
}

\usepackage[dvipsnames]{xcolor}

\begin{document}

\title[Stochastic Schr\"{o}dinger equation for a homodyne measurement setup]{Stochastic Schr\"{o}dinger equation for a homodyne measurement setup of strongly correlated systems}

\author{    Aniket Patra$^1$, Felix Motzoi$^2$ and Klaus M{\o}lmer$^3$    }

\address{$^1$ Center for Theoretical Physics of Complex Systems, Institute for Basic Science (IBS), Daejeon, 34126, Republic of Korea}
\ead{patraaniket@gmail.com}

\address{$^2$ Forschungszentrum J\"{u}lich, Institute of Quantum Control (PGI-8), D-52425 J\"{u}lich, Germany}

\address{$^3$ Department of Physics and Astronomy, University of Copenhagen, Blegdamsvej 17, K{\o}benhavn {\O}, Denmark}

\begin{abstract}
Starting from an experimentally feasible atomic setup, we derive a stochastic Schr\"{o}dinger equation that captures the homodyne detection record of a strongly interacting system. Applying the rotating wave approximation to the linear atom-light coupling, we arrive at a reduced equation formulated solely in terms of atomic operators. In the appropriate limit, this equation converges to that of Gaussian continuous quantum measurement --- revealing that the complexities of real-world detection can, under certain conditions, echo the elegance of idealized theory. To illustrate the utility of this framework, we numerically study the Bose-Hubbard model under continuous observation, showing that time-domain analysis of the measurement signal uncovers rich dynamical features, including quantum jumps, that are obscured in ensemble-averaged spectral data.
\end{abstract}

\begin{indented}\item[]
\end{indented}

\submitto{\jpb}

\maketitle


\section{Introduction}
\label{SecIntro}

Measurements occupy a central position in quantum mechanics \cite{Sakurai}, forming the bridge between theoretical predictions and empirical observation. However, when a quantum system is continuously interrogated by an auxiliary device --- such as a laser probing an atomic ensemble --- the standard framework of unitary Hamiltonian evolution coupled with projective measurements \cite{BlochReview, Greiner2002} becomes inadequate. In such settings, the system under study must be treated as open, with its evolution conditioned on the measurement outcomes. This necessitates a formalism based on weak and continuous measurements, where the measurement record, inherently stochastic, encapsulates the gradual acquisition of information about the system \cite{Wiseman, Belavkin, Percival, Wiseman_PRA, MCWF0, MCWF1, MCWF2, Carmichael, Knight_RMP, SpinHall, PKP_Theory, PKP1, PKP2}. Beyond its foundational interest, the conditioned evolution has practical consequences: it enables feedback control \cite{Feedback_Theory, Feedback_Expt}, offers pathways to suppress decoherence \cite{Milburn_Anti-Decoh}, and facilitates entanglement generation \cite{Milburn_Ent, Cai_Ent}. The link between measurement schemes and stochastic Schr\"{o}dinger equations (SSE) is well established \cite{Wiseman_PRL, JS, JBook, Daley, Nori, NonMarkov1, NonMarkov2, NonMarkov3}, but applying this machinery to strongly correlated many-body systems --- each scheme yielding a distinct evolution --- has only recently opened new paths into nonequilibrium quantum dynamics \cite{ColFric_PRL, RitschReview, N-Queens_Solver, Dicke_Transition, Diss_Phases, MB_Ent_PRL, Memory_NN1, Memory_NN2, Dyn_Frustr, Galitski_SC, Cav_Asstd_Cooling, Monitored_Cooling, Cav_Enh_Trans1, Cav_Enh_Trans2, Cont_Trans_Meas, Meas_Ind_PT1, Meas_Ind_PT2, Daley_NonMarkov, MultMode1, MultMode2, Dicke_Tran, Sherson_PRA, APLBJJ}.

\begin{figure}
\includegraphics[trim={3.3cm 1.5cm 2.9cm 1.5cm},clip, width=1\textwidth]{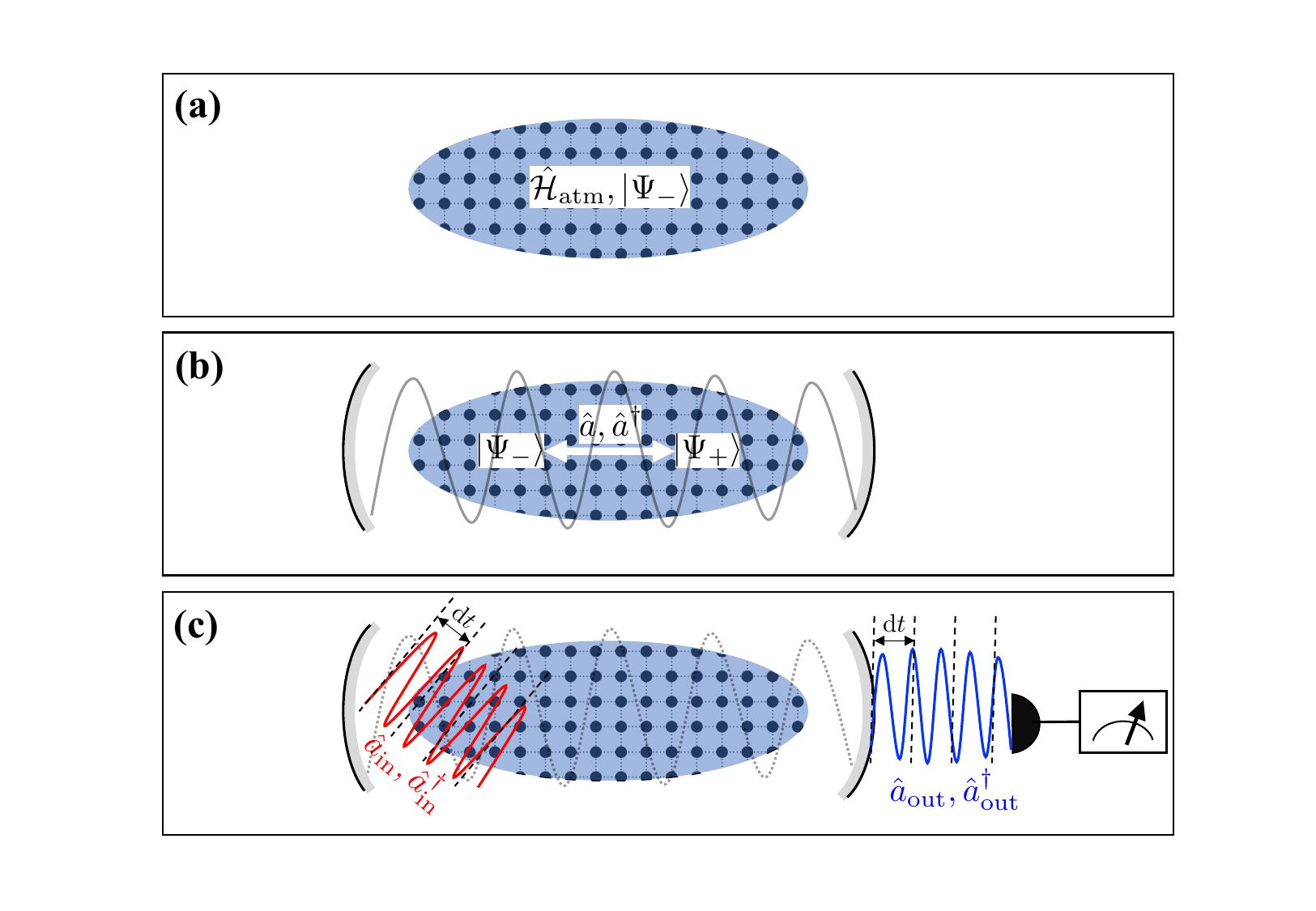}
\caption{Studying strongly correlated systems inside an optical cavity. In the panel \textbf{(a)} of the schematic diagram above, we show the strongly correlated Hamiltonian $\hat{H}_{\text{atm}}$ realized by trapping ultracold atoms in an optical lattice.  The ground state of this Hamiltonian $| \Psi_{-} \rangle$ has interesting properties that one would like to explore. In panel \textbf{(b)}, we depict the strongly correlated atomic system inside a single mode optical cavity.  The annihilation (creation) operator for this mode is $\hat{a} (\hat{a}^\dag)$.  Although the cavity mode photons couple the ground state $| \Psi_{-} \rangle$ to the excited state $| \Psi_{+} \rangle$, it is possible to adiabatically eliminate the excited state and obtain an effective dispersive atom-light Hamiltonian \eqref{disphamiltonian}.  In panel \textbf{(c)}, we show the external driving laser --- input field $(\hat{a}_\mathrm{in}, \hat{a}_\mathrm{in}^\dag)$ --- in red \cite{InOut1, InOut2, InOut3, InOut4, InOut5, InOut6}.  This then gives rise to the output field $(\hat{a}_\mathrm{out}, \hat{a}_\mathrm{out}^\dag)$.  The input field populates the cavity, and the output field --- which is entangled with the strongly correlated system inside the cavity --- is finally detected. The output photons are detected with a homodyne setup, see Fig.\ \ref{HomodynePic}.  In fact, this set up detects the linear combinations of the local oscillator and output mode. We show that in an appropriate limit the measurement record is appropriately described by an SSE \eqref{SSE_Blatt} in terms of only the atomic degrees of freedom, which is equivalent to the SSE that is obtained when one performs a Gaussian continuous quantum measurement of an atomic observable $\hat{\mathcal{M}}_{0}$ on $\hat{\mathcal{H}}_{\text{atm}}$.}
\label{Schematic}
\end{figure}

\begin{figure}
\begin{indented}\item[]
\includegraphics[trim={5.3cm 3cm 5.3cm 2.5cm},clip, scale=0.38]{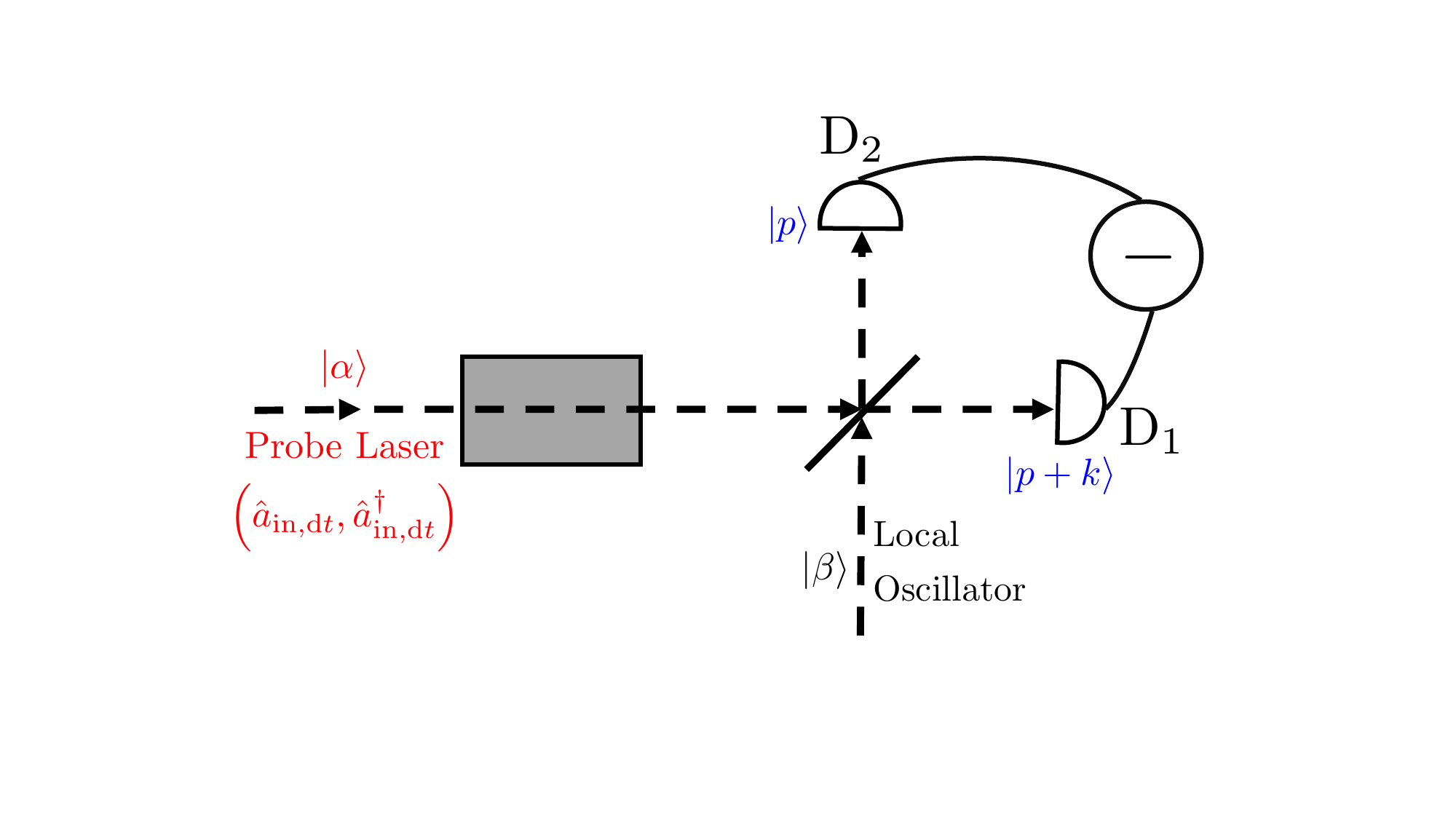}
\caption{In this experimental setup, the gray shaded box represents the cavity, where the strongly interacting Hamiltonian is engineered using cold atoms in optical lattices. The slanted straight line is the 50:50 beam splitter.  The two detectors are marked as $\textrm{D}_1$ and $\textrm{D}_2$. This figure has been reused with modifications from Ref.\ \cite{APLBJJ}.}
\label{HomodynePic}
\end{indented}
\end{figure}

We study a concrete experimental setup, shown in Fig.\ \ref{Schematic}, where a strongly interacting atomic system is embedded in a lossy optical cavity and probed via homodyne detection (Fig.\ \ref{HomodynePic}). Each component --- the atomic ensemble, the driven cavity, the coherent input, and the detection scheme --- is modeled explicitly. The high cavity loss permits adiabatic elimination of the cavity mode, leading to an effective description in terms of atomic operators alone. Continuous monitoring of the output field yields an SSE for the conditioned atomic state. Generally, establishing such a link between a physical measurement scheme and a specific stochastic evolution is nontrivial \cite{Luoma_PRR}. Yet in the limit of weak driving and strong local oscillator, the resulting equation recovers the structure of Gaussian quantum continuous measurement \eqref{GQCM}.

Several works have explored the dynamics of many-body systems under continuous measurement by postulating a SSE from the outset. In particular, Refs.\ \cite{Meas_Ind_PT1, Meas_Ind_PT2, Sherson_PRA, APLBJJ} consider Gaussian quantum continuous measurements of specific Hermitian observables defined in terms of the same degrees of freedom --- typically atomic --- as those appearing in the system Hamiltonian. However, these studies do not derive the SSE from an explicit experimental setup, leaving the physical origin of the measurement scheme implicit. In contrast, Ref.\ \cite{AEBN} begins with a concrete experimental configuration (though distinct from the one considered here) and derives a stochastic master equation. Yet, the resulting dynamics is not expressed solely in terms of the atomic degrees of freedom, nor does the SSE assume the canonical form expected for Gaussian continuous measurements.

A complementary direction emphasizes modeling memory effects and non-Markovian dynamics \cite{NonMarkov1, NonMarkov2, NonMarkov3, Daley_NonMarkov}. While some works, e.g., Ref.\ \cite{Daley_NonMarkov}, obtain SSEs structurally similar to ours, they do not explicitly incorporate the input modes or detection scheme. Others attempt to include these components \cite{NonMarkov2, NonMarkov3}, but do not arrive at a closed-form SSE for the atomic system, largely due to the absence of a separation of timescales that would justify adiabatic elimination.

In this work, we provide a detailed experimental setup that naturally leads to the canonical SSE describing Gaussian quantum continuous measurement of an atomic observable within a strongly correlated many-body system. This construction establishes a concrete physical basis for the class of measurement-induced dynamics previously explored in \cite{Meas_Ind_PT1, Meas_Ind_PT2, APLBJJ}, offering a platform for experimentally accessing rich many-body phenomena driven by continuous quantum observation.

In Sec.\ \ref{SecIntro_to_Measurement}, we review the main paradigms of quantum measurement and establish the terminology used throughout. Sec.\ \ref{SecEffH} introduces the atom-light Hamiltonian for atoms in a Fabry-P\'{e}rot cavity (Fig.\ \ref{Schematic}) and, using input-output theory alongside the Markov approximation \cite{InOut1, InOut2, InOut3, InOut4, InOut5, InOut6}, derives an effective interaction between the atomic observable and the input field, made possible by rapid cavity decay. In Sec.\ \ref{Microscopic_SSE}, we analyze the conditional density matrix in a homodyne detection scheme and show that, in the strong local oscillator limit, the photon number difference introduces a Wiener process. Under additional approximations, the resulting stochastic Schr\"{o}dinger equation reduces to the canonical form associated with Gaussian quantum continuous measurement. Finally, in Sec.\ \ref{Ex_BH}, we numerically investigate this SSE for the Bose-Hubbard model \cite{BH}, demonstrating that time-dependent observables reveal the superfluid–Mott insulator transition and showcasing quantum trajectories in the Zeno regime \cite{zenoECG, zenophase, klaus2, zenoSDG}.
\\

\section{Quantum Measurements: Setting the Stage and the Symbols} \label{SecIntro_to_Measurement}

In this section, we sketch the main frameworks for quantum measurement and fix the terminology that anchors the rest of this work. In a typical quantum mechanics text book, we associate an observable $\hat{\Lambda}$, which is generally represented as a Hermitian matrix, to the measured physical property, e.g., momentum. It is then possible to write $\hat{\Lambda} = \sum_{\lambda} \lambda \hat{\Pi}_{\lambda}$, where $\lbrace \lambda \rbrace$ are the real eigenvalues of $\hat{\Lambda}$ and $\hat{\Pi}_{\lambda} = | \lambda \rangle \langle \lambda |$ are rank-1 projectors.  If the projective measurement produces a result $\lambda$, the quantum state must have been projected by $\hat{\Pi}_{\lambda}$ into the eigenstate $| \lambda \rangle$. For instance, the time of flight measurement that one performs to probe different phases of interacting atomic Hamiltonians in optical lattices are analyzed within this projective measurement paradigm.

In this paper, however, we focus on a more realistic description of an experimental setup, where the system of interest is not directly measured. Rather, the system interacts with a second quantum system (called the \textit{meter} or the \textit{apparatus}), and the meter is later measured projectively.  Instead of using a projective measurement description, such systems are better described in terms of a generalized measurement. Here one obtains a state that is peaked about an eigenstate $| \lambda \rangle$ with a certain width. A large width means more uncertainty in obtaining the corresponding eigenvalue $\lambda$.  A generalized measurement, where one manages to extract only infinitesimal information about the system, is referred to as a \textit{weak measurement}. 

One need not explicitly include the meter in the description of a weak measurement. It is possible to describe the measurement scheme by the action of a set of generalized measurement operators $\lbrace \hat{M}_{r} \rbrace$ on the system, where $\lbrace r \rbrace$ denotes the corresponding measurement outcomes.  Note that, by $\lbrace r \rbrace$, we denote the real eigenvalues of an ancillary observable.  After the measurement process, if one obtains a result $r$, the state of the system is given by $\hat{M}_{r}| \Psi \rangle/\mathcal{P}_{r},$ where $\mathcal{P}_{r}$ is the probability of getting the particular result $r$. This probability $\mathcal{P}_{r}$ is defined in terms of the \textit{probability operators} or \textit{effects} $\hat{E}_{r} = \hat{M}_{r}^\dag \hat{M}_{r}$ as $\mathcal{P}_{r} = \langle \Psi |\hat{E}_{r} | \Psi \rangle.$ Since $\sum_{r} \mathcal{P}_{r} = 1$, one must have $\sum_{r} \hat{E}_{r} = \mathbb{1}$. Compare this with the resolution of identity $\sum_{\lambda} \hat{\Pi}_{\lambda} = \mathbb{1}$ for projective measurements. The set of all these effects $\lbrace \hat{E}_{r}: r \rbrace$ is referred to as a \textit{positive operator-valued measure} (POVM).

The generalized measurement operator given by
\begin{equation}
\label{GQCM}
\hat{M}(\alpha) = \left(\frac{4\mathcal{K}\mathrm{d} t}{\pi}\right)^{1/4}\int_{-\infty}^{+\infty} e^{-2\mathcal{K}\mathrm{d} t \left(x - \alpha\right)^{2}}\left | x \right \rangle \left \langle x \right | \mathrm{d} x
\end{equation}
corresponds to the \textit{Gaussian quantum continuous measurement} of the Hermitian operator $\hat{X}$ at a sequence of intervals of length $\mathrm{d} t$ and with measurement strength $\mathcal{K}$ \cite{JS, JBook}. The Hermitian operator $\hat{X}$ has a continuous spectrum of eigenvalues $x$ and orthonormal eigenstates $|x\rangle$ with the property $\langle x| x^{\prime}\rangle = \delta \left(x - x^{\prime}\right)$. The continuous indices $\alpha$ are the eigenvalues of an ancillary observable and are the possible measurement outcomes. They are not, however, the eigenvalues of either $\hat{M}(\alpha)$ or $\hat{X}$. This work establishes the link between the experimental architecture of Ref.\ \cite{APLBJJ} and the theoretical framework of Gaussian continuous quantum measurement.
\\

\section{Derivation of the Effective Hamiltonian}
\label{SecEffH}

We start with the atom-light Hamiltonian of Fig.\ \ref{Schematic}(b), where the atomic ground and excited states are coupled via the cavity mode. Assuming that the detuning between the laser frequency and the frequency of the atomic excitation is large, we adiabatically eliminate the atomic excited state. This leads to the dispersive atom-light interacting Hamiltonian of the cavity. 

In the dipole approximation, the full atom-light Hamiltonian is given by
\begin{multline}
\hat{\mathcal{H}}=\hat{\mathcal{H}}_A+\hat{\mathcal{H}}_{\text{atm}}+\hat{\mathcal{H}}_a+g\hat{a}\int\hat{\Psi}_+^\dag(x)f_a(x,\omega_L)\hat{\Psi}_-(x)\;\mathrm{d}x\\
+g\hat{a}^\dag\int\hat{\Psi}_-^\dag(x)f_a(x,\omega_L)\hat{\Psi}_+(x)\;\mathrm{d}x,
\label{HamPanelB}
\end{multline}
where $\hat{\Psi}_{+(-)}$ is the bosonic field operator for the excited (ground) state of the atoms and we have neglected the fast-rotating terms. The probe field, which we assume to be in resonance with the cavity, has frequency $\omega_L$ and $g$ is the coupling strength \cite{BlochReview}. Moreover, we have $\hat{\mathcal{H}}_A=\omega_{eg}\int \text{d} x \;\hat{\Psi}_+^\dag(x)\hat{\Psi}_+(x)$ corresponding to the atomic excited state and $\hat{\mathcal{H}}_a=\omega_L\hat{a}^\dag\hat{a}$ corresponding to the free Hamiltonian of the cavity light field with the spatial mode function $f_a(x,\omega_L)$. Here $\hat{\mathcal{H}}_{\text{atm}}$ is expressed in terms of the ground-state bosonic field operators $\hat{\Psi}_{-}$ and $\hat{\Psi}_{-}^{\dag}$. For a concrete example, where $\hat{\mathcal{H}}_{\text{atm}}$ is the same as the Bose-Hubbard Hamiltonian, see Ref.~\cite{BH}. 

We introduce the fields that oscillate slowly with time
\begin{equation}
\hat{\widetilde{a}} = e^{i\omega_Lt}\hat{a},  \quad \hat{\widetilde{\Psi}}_+ = \hat{\Psi}_+e^{i\omega_{eg}t}, \quad \hat{\widetilde{\Psi}}_- \equiv \hat{\Psi},
\label{NewFields}
\end{equation}
where $\hat{\Psi}_{-}$ is assumed to be already slowly oscillating, and we have included the free (comparatively fast) evolution of the excited state and the light field into the new field operators (in the Schr\"{o}dinger picture itself). Recall that in the Heisenberg picture an observable satisfies the equation
\begin{equation}
\label{HeisenbergEOM}
\frac{\mathrm{d} }{\mathrm{d} t}\hat{A}_{H}(t) = i \left[ \hat{\mathcal{H}}_{H},  \hat{A}_{H}(t)\right] + \left( \frac{\partial \hat{A}_{S}}{\partial t} \right)_{H},
\end{equation}
where the subscripts $H$ and $S$ denote Heisenberg and Schr\"{o}dinger picture, respectively. Using this, we find that the new fields have the following equations of motion:
\begin{subequations}
\begin{align}
&\frac{\mathrm{d} \hat{\widetilde{a}}}{\mathrm{d} t}=-ig\int\hat{\widetilde{\Psi}}_-^\dag(x)f_a(x,\omega_L)\hat{\widetilde{\Psi}}_+(x)e^{i\Delta t}\mathrm{d}x,\label{adela}\\
&\frac{\mathrm{d} }{\mathrm{d} t}\hat{\widetilde{\Psi}}_+(x) = -ig\hat{\widetilde{a}}\hat{\widetilde{\Psi}}_-(x)f_a(x,\omega_L)e^{-i\Delta t},
\label{adelpsi+}
\end{align}
\end{subequations}
where $\Delta=\omega_L-\omega_{eg}$ is the detuning of the probe from the atomic transition. We assume $\lvert\Delta\rvert\gg g,$ and initially all the atoms are in their ground state.  In that case, by integrating Eq.\ \eqref{adelpsi+} by parts, we obtain
\begin{equation}
\hat{\widetilde{\Psi}}_+(x) = \frac{g}{\Delta}\hat{\widetilde{a}}\hat{\widetilde{\Psi}}_-(x)f_a(x,\omega_L)e^{-i\Delta t} - \frac{gf_a(x,\omega_L)}{\Delta}\int \left[\frac{\mathrm{d}}{\mathrm{d}t}\left(      \hat{\widetilde{a}}\hat{\widetilde{\Psi}}_-(x)     \right)   e^{-i\Delta t} \right]\mathrm{d}t.
\label{adelpsi+1}
\end{equation} 
Since, in the second term of Eq.\ \eqref{adelpsi+1}, a slowly varying function $\mathrm{d}[\hat{\widetilde{a}}\hat{\widetilde{\Psi}}_-(x)]/\mathrm{d}t$ multiplies the fast oscillating exponential $e^{-i\Delta t}$, the integral  vanishes due to the Riemann-Lebesgue lemma. As a result, we obtain
\begin{equation}
\hat{\widetilde{\Psi}}_+(x) = \frac{g}{\Delta}\hat{\widetilde{a}}\hat{\widetilde{\Psi}}_-(x)f_a(x,\omega_L)e^{-i\Delta t},
\end{equation}
which allows us to adiabatically eliminate the excited atomic state (see also Ref.\ \cite{BH_PRA}) and write
\begin{equation}
\label{suppBHEOM}
\frac{\mathrm{d} \hat{\widetilde{a}}}{\mathrm{d} t} = -\frac{ig^2}{\Delta}\hat{\widetilde{a}}\int|f_a(x,\omega_L)|^2\hat{\widetilde{\Psi}}_-^\dag(x)\hat{\widetilde{\Psi}}_-(x)\;\mathrm{d}x.
\end{equation}
With the help of Eq.\ \eqref{HeisenbergEOM}, one notices that the effective Hamiltonian
\begin{equation}
\label{disphamiltonianNew}
\hat{\widetilde{\mathcal{H}}} = \hat{\mathcal{H}}_{\text{atm}}+\frac{g^2}{\Delta}\int|f_a(x,\omega_L)|^2\hat{\widetilde{\Psi}}^\dag(x)\hat{\widetilde{\Psi}}(x)\hat{\widetilde{a}}^\dag\hat{\widetilde{a}}\;\mathrm{d}x
\end{equation}
gives rise to the equation of motion (\ref{suppBHEOM}), where the subscript for the atomic field operator has now been dropped. This then gives the following effective Hamiltonian in terms of the original field operators:
\begin{equation}\label{disphamiltonian}
\hat{\mathcal{H}}=\hat{\mathcal{H}}_{\text{atm}} + \hat{\mathcal{H}}_a
+ \frac{g^2}{\Delta}\int|f_a(x,\omega_L)|^2\hat{\Psi}^\dag(x)\hat{\Psi}(x)\hat{a}^\dag\hat{a}\;\mathrm{d}x.
\end{equation}

The localized (at the minima of the optical lattice potential) Wannier functions form a complete set for the first band of the atomic system. We now expand the field operators in terms of these functions as
\begin{equation}
\hat{\Psi}(x)=\sum_j\hat{b}_{j}w_j(x).
\label{Wannier_Basis}
\end{equation}
The creation operators $\hat{b}^{\dagger}_{i}$ creates  a boson at the lattice site $i$. The relevant matrix determining the spatial influence of the measurement is obtained by expressing the integral in Hamiltonian (\ref{disphamiltonian}) in terms of these Wannier functions. The resulting Hamiltonian reads
\begin{align}\label{ham1}
\hat{\mathcal{H}}=\hat{\mathcal{H}}_{\text{atm}} + \hat{\mathcal{H}}_a +\sum_{j,k}M_{jk}\hat{b}_j^\dag\hat{b}_k\hat{a}^\dag\hat{a},
\end{align}
with
\begin{equation}
\label{MInt}
M_{jk}=\frac{g^2}{\Delta}\int|f_a(x,\omega_L)|^2w_j^*(x)w_k(x)\mathrm{d}x.
\end{equation}

The cavity mode $(\hat{a}, \hat{a}^\dag)$ in the Hamiltonian (\ref{ham1}) is populated by an external laser as depicted in Fig.\ \ref{Schematic}(c). Using the input-output theory to model this system \cite{InOut1, InOut2, InOut3, InOut4, InOut5, InOut6}, we obtain the Heisenberg-Langevin equation
\begin{align}
\dot{\hat{a}}= -i\omega_{L}\hat{a}-i\sum_{jk}M_{jk}\hat{b}^\dag_j\hat{b}_k\hat{a}-\kappa\hat{a}+\eta+\sqrt{2\kappa}\hat{a}_\mathrm{in},
\end{align}
with $\hat{a}_\mathrm{in}$ being the annihilation operator corresponding to the input field, $2\kappa$ the cavity mode decay rate, and $\eta$ the pump power \cite{JBook}. We displace $\hat{a}$ by its coherent expectation value
\begin{equation}
 \hat{a}\to\mathcal{A}+\hat{a}
\end{equation}
such that we have
\begin{multline}
\dot{\hat{a}}+\dot{\mathcal{A}}=-i\omega_{L}\mathcal{A}-i\omega_{L}\hat{a}-i\sum_{jk}M_{jk}\hat{b}^\dag_j\hat{b}_k \mathcal{A} \\
-i\sum_{jk}M_{jk}\hat{b}^\dag_j\hat{b}_k\hat{a}-\kappa\mathcal{A}-\kappa\hat{a}+\eta+\sqrt{2\kappa}\hat{a}_\mathrm{in}.
\label{eq:SuppCoh1}
\end{multline}
We choose $\mathcal{A} = \mathcal{A}_0$ to be the solution of
\begin{align}
\dot{\mathcal{A}}=-i\omega_{L}\mathcal{A}-\kappa\mathcal{A}+\eta,
\label{eq:SuppCoh2}
\end{align}
such that $\hat{a}$ obeys
\begin{align}
\dot{\hat{a}}=-i\omega_{L}\hat{a}-i\mathcal{A}_0\sum_{jk}M_{jk}\hat{b}^\dag_j\hat{b}_k-\kappa\hat{a}+\sqrt{2\kappa}\hat{a}_\mathrm{in},
\label{IntraCavityEOM}
\end{align}
where we have neglected the term coupling the quantum operators directly compared to the term enhanced by $\mathcal{A}_0$. This equation then points to the following effective coupling Hamiltonian:
\begin{subequations}
\label{FinAtomCavEffH}
\begin{align}
\hat{\mathcal{H}} =& \hat{\mathcal{H}}_{\text{atm}}+\hat{\mathcal{H}}_a+\hat{\mathcal{M}}(\mathcal{A}^{*}_0 \hat{a} +\mathcal{A}_0 \hat{a}^\dag), \label{aNew} \\
\hat{\mathcal{M}} =& \sum_{j,k}M_{jk}\hat{b}_j^\dag\hat{b}_k.
\end{align}
\end{subequations}

Since the input mode (before entering the cavity) and the atomic degrees of freedom are decoupled, we would like to obtain a Hamiltonian in terms of the input field $(\hat{a}_\mathrm{in}, \hat{a}_\mathrm{in}^\dag)$ instead of the intra-cavity field $(\hat{a}, \hat{a}^\dag)$.  To that end,  we rewrite Eq.\ \eqref{IntraCavityEOM} as
\begin{align}
\frac{    \mathrm{d} \hat{\widetilde{a}}    }{    \mathrm{d} t   }    =     - i e^{i\omega_Lt} \mathcal{A}_0\sum_{jk}M_{jk}\hat{b}^\dag_j\hat{b}_k    -    \kappa \hat{\widetilde{a}}    +    \sqrt{2\kappa} e^{i\omega_Lt} \hat{a}_\mathrm{in},
\label{IntraCavityEOM_atilde}
\end{align}   
where $\hat{\widetilde{a}} = e^{i\omega_Lt}\hat{a}$.  In the bad cavity limit, utilizing the fast relaxation of $\hat{\widetilde{a}}$ and equating $\dot{\hat{\widetilde{a}}}$ to zero in Eq.\ \eqref{IntraCavityEOM_atilde}, we obtain
\begin{equation}
\hat{a} = \frac{1}{\kappa}    \left(  -i\mathcal{A}_0\hat{\mathcal{M}} + \sqrt{2\kappa}\hat{a}_\mathrm{in}  \right)
\label{IntraCavityAdElm}
\end{equation}  
Substituting above into Eq.\ \eqref{aNew}, we obtain the following effective coupling Hamiltonian:
\begin{multline}
\hat{\mathcal{H}} = \hat{\mathcal{H}}_{\text{atm}}     +     \frac{  \left| \mathcal{A}_0 \right|^{2} \omega_{L}  }{  \kappa^{2}  }    \hat{\mathcal{M}}^{2}  +    \frac{  2 \omega_{L}  }{  \kappa  }    \hat{a}^{\dag}_\mathrm{in}  \hat{a}_\mathrm{in}           \\    
+  \sqrt{    \frac{  2  }{  \kappa  }    } \hat{\mathcal{M}} \left[    \mathcal{A}^{*}_0  \left(  1 + \frac{  i\omega_{L}  }{  \kappa  }  \right) \hat{a}_\mathrm{in}    +    \mathcal{A}_0 \left(  1 - \frac{  i\omega_{L}  }{  \kappa  }  \right)  \hat{a}^{\dag}_\mathrm{in}     \right].   
\label{EffHInput} 
\end{multline}    
The output mode (corresponding to the light coming out of the cavity) is related to the input and the intra-cavity mode by the relation \cite{InOut1, InOut2, InOut3, InOut4, InOut5, InOut6}
\begin{equation}
\hat{a}_\mathrm{out}(t) - \hat{a}_\mathrm{in}(t) = \sqrt{2\kappa} \hat{a}.
\label{InOutEq}
\end{equation}
However, since the output mode is entangled with the atomic degrees of freedom, we resist  eliminating the input mode further in Eq.\ \eqref{EffHInput} and obtaining an effective Hamiltonian in terms of the output mode. 

We now note that the creation and annihilation operators corresponding to the input mode obey the commutation relation 
\begin{equation}
\left[  \hat{a}_\mathrm{in}(t),  \hat{a}^{\dag}_\mathrm{in}(t^{\prime})  \right] = \delta(  t - t^{\prime}  ), 
\label{InputComm}
\end{equation} 
which is not the same one obeyed by the intra-cavity mode operators. The latter obey
\begin{equation}
\left[  \hat{a}(t),  \hat{a}^{\dag}(t^{\prime})  \right] = \mathbb{1}.
\label{IntraComm}
\end{equation} 
Therefore, we define new coarse-grained  operators as
\begin{subequations}
\begin{align}
&\hat{a}_{  \mathrm{in},  \mathrm{coarse}  }(t) = \frac{1}{\sqrt{\mathrm{d} t}}    \int_{t}^{t + \mathrm{d} t}    \hat{a}_\mathrm{in}(t)  \mathrm{d} t \approx \hat{a}_\mathrm{in}(t)  \sqrt{\mathrm{d} t},     \\
&\hat{a}^{\dag}_{  \mathrm{in},  \mathrm{coarse}  }(t) = \frac{1}{\sqrt{\mathrm{d} t}}    \int_{t}^{t + \mathrm{d} t}    \hat{a}^{\dag}_\mathrm{in}(t)  \mathrm{d} t \approx \hat{a}^{\dag}_\mathrm{in}(t)  \sqrt{\mathrm{d} t}.
\end{align}
\label{CInput}
\end{subequations}
While defining these new operators, we assume that the operators $\hat{a}_\mathrm{in}(t)$ and $\hat{a}^{\dag}_\mathrm{in}(t)$ remain constant inside the time interval $(t, t+\mathrm{d} t]$.  

This is justified because the frequency of the photons are controlled by the parameters $(\ll 1/\mathrm{d} t)$ of the Hamiltonian. As a result, the higher frequency modes $(\sim 1/\mathrm{d} t)$ of $\hat{a}_\mathrm{in}(t)$ remain unoccupied \cite{AEBN}.  Putting Eqs.\ \eqref{InputComm} and \eqref{CInput} together we observe that the operators $\hat{a}_{  \mathrm{in},  \mathrm{coarse}  }(t)$ and $\hat{a}^{\dag}_{  \mathrm{in},  \mathrm{coarse}  }(t)$ obey the same commutation relation as in Eq.\ \eqref{IntraComm}.  Finally, we write the effective Hamiltonian as
\begin{equation}
\hat{\mathcal{H}} = \hat{\widetilde{\mathcal{H}}}_{\text{atm}}     +     \alpha_{0}\hat{\mathcal{M}} \left(  \hat{a}_{  \mathrm{in},  \mathrm{d} t  }    +    \hat{a}^{\dag}_{  \mathrm{in},  \mathrm{d} t  }     \right),
\label{EffHInputFin} 
\end{equation} 
where we use the following definitions:
\begin{subequations}
\begin{align}
\alpha_{0}    e^{i\varphi}    &\equiv    \mathcal{A}_0   \sqrt{\frac{2}{\kappa\mathrm{d} t}} \left(  1 - \frac{  i\omega_{L}  }{  \kappa  }  \right), \label{EffHInputFinDefn1}    \\
\hat{a}_{  \mathrm{in},  \mathrm{d} t  }(t)    &\equiv    \hat{a}_{  \mathrm{in},  \mathrm{coarse}  }(t)  e^{  -i\varphi  -i\widetilde{ \omega }_L t  },     \label{EffHInputFinDefn2}    \\ 
\widetilde{ \omega }_L    &\equiv    \frac{  2  \omega_{L}  }{  \kappa \mathrm{d} t  },    \label{EffHInputFinDefn3}    \\        
\hat{\widetilde{\mathcal{H}}}_{\text{atm}}    &\equiv     \hat{\mathcal{H}}_{\text{atm}}     +     \frac{  \left| \mathcal{A}_0 \right|^{2} \omega_{L}  }{  \kappa^{2}  }    \hat{\mathcal{M}}^{2}.    \label{EffHInputFinDefn4}
\end{align}
\end{subequations}
In this Hamiltonian, we do not include the term proportional to $\hat{a}^{\dag}_\mathrm{in}  \hat{a}_\mathrm{in}$ from Eq.\ \eqref{EffHInput} since we have included the time evolution due to this part in the definition of the operator $\hat{a}_{  \mathrm{in},  \mathrm{d} t  }(t)$.

In this section, we begin with the atom–light Hamiltonian, where the atomic excited state is dipole-coupled to the ground state via the cavity mode, which itself is driven by an external laser. The coupling between the cavity and the input field is modeled using input-output theory in the Markovian limit \cite{InOut1, InOut2, InOut3, InOut4, InOut5, InOut6}. In the regime of large detuning, we adiabatically eliminate the excited atomic state, and under the bad cavity condition, the cavity mode is similarly eliminated. This yields an effective Hamiltonian \eqref{EffHInputFin}, expressed in terms of the atomic ground-state field operators and the input modes. Expanding the atomic fields in the Wannier basis \eqref{Wannier_Basis} of a chosen optical lattice \eqref{OptLattBH} reduces the system to a Bose-Hubbard Hamiltonian with additional corrections. The resulting effective Hamiltonian is then used in Sec.\ \ref{Microscopic_SSE} to describe the unitary evolution of the joint atom–input laser system, which is initially assumed to be in a product state prior to their interaction.

\section{Stochastic Schr\"{o}dinger Equation from the Microscopic Hamiltonian}
\label{Microscopic_SSE}

The Fabry-P\'erot cavity that contains the atomic system produces a standing wave of frequency $\omega_\mathrm{L}$ and gives rise to the spatial mode function $f_a(x,\omega_\mathrm{L})$ in Hamiltonian \eqref{HamPanelB}.  As was explained in the section before, in the Hamiltonian \eqref{EffHInputFin}, we have expressed the intra-cavity photon operators $\left(\hat{a}, \hat{a}^\dag\right)$ in terms of the input field operators $\left(\hat{a}_{  \mathrm{in},  \mathrm{d} t  }, \hat{a}_{  \mathrm{in},  \mathrm{d} t  }^{\dag}\right)$ and the atomic operators by using adiabatic elimination.    This has, in effect, eliminated the intra-cavity mode from the subsequent equations. The goal of this section is to start from the Hamiltonian (\ref{EffHInputFin}) and derive the corresponding SSE, when one implements the weak measurement scheme.  This SSE, in the appropriate limit, becomes the one corresponding to a Gaussian continuous measurement \cite{JBook, JS, APLBJJ}.

\subsection{Origin of Wiener Noise: Measurement of $\hat{P}$ Quadrature of the Output}
\label{Sec_PQuadrature}

Before deriving the SSE for our experimental setup, we show that it is possible to obtain a continuous SSE with an additive Wiener noise if one is able to measure the $\hat{P}$ quadrature of the output field.  We  start with the product state of the atomic part and the input field $\left|\Phi(t)\right\rangle = \left|\Psi(t)\right\rangle  \left|\alpha\right\rangle$, where the atomic wavefunction is $\left|\Psi(t)\right\rangle$.  For future use, we expand the atomic state $\left|\Psi\right\rangle$ in the eigenbasis of $\hat{\mathcal{M}}$ --- which is denoted as $\left\lbrace\left|\psi_{m}\right\rangle \right\rbrace$ --- and write
\begin{equation}
\left|\Psi\right\rangle = \sum_{m}C_{m}\left|\psi_{m}\right\rangle.
\label{EigExp}
\end{equation}

We assume that at time $t$ the coarse-grained modes (corresponding to the time-segment $\text{d}t \ll 1$) of the input laser can be described by the coherent states $\left|\alpha\right\rangle$, where $\hat{a}_{  \mathrm{in},  \mathrm{d} t  }    \left|\alpha\right\rangle  = \alpha  \left|\alpha\right\rangle$.  We also assume that $\alpha$ is real.  A coherent state $\left|\zeta\right\rangle$ is written as a displacement operator $\hat{D}(\zeta) = \exp\left(\zeta\hat{a}^\dag - \zeta^*\hat{a}\right)$ acting on the vacuum state, where $\zeta$ can be complex in general \cite{Knight}. The displacement operators have the property
\begin{equation}
\hat{D}(\zeta)\hat{D}(\eta) = e^{\left(\zeta\eta^* - \zeta^*\eta\right)/2}\hat{D}(\zeta + \eta),
\label{DispOp}
\end{equation}
which is proven using the Baker-Campbell-Hausdorff formula.  

Making use of Eqs.\ \eqref{DispOp} and \eqref{EigExp},  we then obtain the entangled wavefunction for the atomic system and the output field as:
\begin{gather}
\begin{aligned}
& \left|\Phi(t + \mathrm{d} t)\right\rangle    \\
&=    e^{  -i\hat{\widetilde{\mathcal{H}}}_{\text{atm}}\mathrm{d} t  }    e^{  -i\alpha_0\hat{\mathcal{M}}\left(  \hat{a}_{  \mathrm{in},  \mathrm{d} t  }    +    \hat{a}^{\dag}_{  \mathrm{in},  \mathrm{d} t  }     \right)  \mathrm{d} t  }\left|\Psi\right\rangle\left|\alpha\right\rangle      \\
&=    e^{  -i\hat{\widetilde{\mathcal{H}}}_{\text{atm}}\mathrm{d} t  }    e^{  -i\alpha_0\hat{\mathcal{M}}\left(  \hat{a}_{  \mathrm{in},  \mathrm{d} t  }    +    \hat{a}^{\dag}_{  \mathrm{in},  \mathrm{d} t  }     \right)  \mathrm{d} t  }    \sum_{m}C_{m}\left|\psi_{m}\right\rangle    \left|\alpha\right\rangle    \\
&=     e^{  -i\hat{\widetilde{\mathcal{H}}}_{\text{atm}}\mathrm{d} t  }   \sum_{m}C_{m} \left[    \underbrace{e^{  (-i\alpha_0 m \mathrm{d} t)\hat{a}^{\dag}_{  \mathrm{in},  \mathrm{d} t  } - (-i\alpha_0 m \mathrm{d} t)^{*}  \hat{a}_{  \mathrm{in},  \mathrm{d} t  }  }\left|\alpha\right\rangle}_{= \hat{D}(-i\alpha_0 m \mathrm{d} t)\hat{D}(\alpha)\left|0\right\rangle}    \right]    \\
&=    e^{  -i\hat{\widetilde{\mathcal{H}}}_{\text{atm}}\mathrm{d} t  }    \sum_{m}C_{m}e^{-i\alpha_0\alpha m \mathrm{d} t}\left|\psi_{m}\right\rangle    \left|\alpha - i\alpha_0 m \mathrm{d} t\right\rangle    \\
&=    e^{  -i    \left(    \hat{\widetilde{\mathcal{H}}}_{\text{atm}}    +     \alpha_0\alpha  \hat{\mathcal{M}}     \right)    \mathrm{d} t  }    \sum_{m}C_{m}    \left|\psi_{m}\right\rangle    \left|\alpha - i\alpha_0 m \mathrm{d} t\right\rangle.
\end{aligned}
\label{StateBeforeBS}
\end{gather}
Note that the coherent state $ \left|\alpha - i\alpha_0 m \mathrm{d} t\right\rangle$ here corresponds to the output light.  The entanglement between the atomic system in the cavity and the output field is evident from the expression in the last line of Eq.\ \eqref{StateBeforeBS}.

We expand the output coherent state further in the continuum basis $\lbrace \left|\mathcal{P}\right\rangle \rbrace$ of the momentum quadrature as
\begin{equation}
\left|\Phi(t + \mathrm{d} t)\right\rangle    =   \int \mathrm{d} \mathcal{P} \;\;\;    e^{  -i    \left(    \hat{\widetilde{\mathcal{H}}}_{\text{atm}}    +     \alpha_0\alpha  \hat{\mathcal{M}}     \right)        \mathrm{d} t  }    \sum_{m}C_{m}    \left|\psi_{m}\right\rangle    \left\langle\mathcal{P}\right|    \left.\alpha - i\alpha_0 m \mathrm{d} t\right\rangle    \left|\mathcal{P}\right\rangle. 
 \label{StateBeforeBSMomBasis}
\end{equation} 
Here we apply the notation of Ref.\ \cite{Wiseman} to write
\begin{equation}
 \hat{a}_{  \mathrm{in},  \mathrm{d} t  }    =    \frac{1}{\sqrt{2}}  \left(  \frac{\hat{Q}}{\sigma}  +  i \sigma \hat{P}  \right),
\end{equation}
where $\hat{Q}$ is the position quadrature, $\hat{P}$ is the momentum quadrature, and $\sigma$ depends on the details of the quadrature measurement. We have also equated $\hbar$ and the mass parameter to one in the simple harmonic oscillator Hamiltonian.  

From the wavefunction \eqref{StateBeforeBSMomBasis}, given the momentum quadrature measurement outcome $\mathcal{P}$, one can write the conditional (a-posteriori) atomic state as
\begin{equation}
\left|\Psi_{\mathcal{P}}(t + \mathrm{d} t)\right\rangle    =    e^{  -i    \left(    \hat{\widetilde{\mathcal{H}}}_{\text{atm}}    +     \alpha_0\alpha  \hat{\mathcal{M}}     \right)        \mathrm{d} t  }    \sum_{m}C_{m}    \Xi_{\widetilde{\alpha}_m}(\mathcal{P})    \left|\psi_{m}\right\rangle,
 \label{CondAtmStateBeforeBS}
\end{equation}  
where the momentum probability amplitude $\Xi_{\widetilde{\alpha}_m}(\mathcal{P})$ is written as 
\begin{equation}
\Xi_{\widetilde{\alpha}_m}(\mathcal{P})    =    \left(  \pi/\sigma^2  \right)^{-1/4}  \exp  \left[  - iq_{m0}\mathcal{P} - \left(  \mathcal{P}  -  p_{m0}  \right)^{2}\sigma^{2}/2   \right],
\label{MomProbAmpl}
\end{equation} 
with $\widetilde{\alpha}_m, q_{m0}$ and $p_{m0}$ defined as 
\begin{subequations}
\begin{align}
\widetilde{\alpha}_m    &\equiv    \alpha - i\alpha_0 m \mathrm{d} t,  \\
q_{m0}    &=     \left\langle \widetilde{\alpha}_m\right|  \left.   \hat{Q}  \right|  \left.   \widetilde{\alpha}_m  \right\rangle  = \sqrt{2}\sigma \mathrm{Re}\left[ \widetilde{\alpha}_m  \right]  = \sqrt{2} \sigma \alpha,   \label{MomQuadDefn1} 
\end{align}
\begin{align}
p_{m0}   &=     \left\langle \widetilde{\alpha}_m\right|  \left.   \hat{P}  \right|  \left.   \widetilde{\alpha}_m  \right\rangle  = \frac{\sqrt{2}}{\sigma} \mathrm{Im}\left[ \widetilde{\alpha}_m  \right]  = -\frac{  \sqrt{2}\alpha_0 m \mathrm{d} t  }{\sigma}. \label{tMomQuadDefn2}   
\end{align}
\label{MomProbAmplDefn}
\end{subequations}

From the conditional state \eqref{CondAtmStateBeforeBS}, we obtain the probability of the measurement outcome $\mathcal{P}$, by taking a trace over the atomic degree of freedom, as
\begin{equation}
\label{MomQuadProb}
\begin{split}
P(\mathcal{P})   &=   \sum_{m}  \left|C_{m}\right|^{2}  \left|\Xi_{\widetilde{\alpha}_m}(\mathcal{P})\right|^{2}  \\
&=    \sum_{m}  \left|C_{m}\right|^{2}    \sqrt{\frac{\sigma^2}{\pi}}  \exp  \left[  - \left(  \mathcal{P}  +  \frac{  \sqrt{2}\alpha_0 m \mathrm{d} t  }{\sigma}  \right)^{2}\sigma^{2}   \right] \\
& \overset{*}{\approx}    \sum_{m}  \left|C_{m}\right|^{2}    \sqrt{\frac{\sigma^2}{\pi}}     e^{  -\sigma^2  \mathcal{P}^2  }   \exp   \left(   -  2\sqrt{2}\mathcal{P}\sigma   \alpha_0     m     \mathrm{d} t \right)    \\
& \overset{*}{\approx}    \sum_{m}  \left|C_{m}\right|^{2}    \sqrt{\frac{\sigma^2}{\pi}}     e^{  -\sigma^2  \mathcal{P}^2  }    \left[  1  -  2\sqrt{2}\mathcal{P}\sigma   \alpha_0     m     \mathrm{d} t \right]    \\
&=     \sum_{m}  \left|C_{m}\right|^{2}    \sqrt{\frac{\sigma^2}{\pi}}     e^{  -\sigma^2  \mathcal{P}^2  }    \left[  1  -  2\sqrt{2}\mathcal{P}\sigma   \alpha_0    \left\langle \hat{\mathcal{M}} \right\rangle     \mathrm{d} t \right]    \\
& \overset{*}{\approx}     \exp  \left[   - \sigma^{2}    \left(  \mathcal{P}  -  \bar{p}_{0}  \right)^{2}   \right],      
\end{split}
\end{equation}
where 
\begin{equation}
\bar{p}_{0}      =     \sqrt{2}\alpha_0     \left\langle \hat{\mathcal{M}} \right\rangle     \mathrm{d} t  /  \sigma. 
\end{equation}
In the third, fourth and the last line (indicated by $*$) of Eq.\ \eqref{MomQuadProb}, we have neglected $\mathcal{O}(\mathrm{d} t^2)$ terms. 

We therefore conclude from Eq.\ \eqref{MomQuadProb} that a new variable
\begin{equation}
\mathrm{d} W = \left(  2 \sigma \mathrm{d} t    \right)^{1/2}    \left[  \hat{\mathcal{P}}  -  \bar{p}_{0}    \right] ,
\label{MomQuadWienerDefn}
\end{equation}
will have a Gaussian probability distribution with mean zero and variance $\mathrm{d} t$ and can be interpreted as the Wiener process. Defining $\mathcal{K}_{\mathcal{P}} \equiv 1/(4 \sigma^2 \mathrm{d} t)$, we obtain the stochastic variable
\begin{equation}
\hat{\mathcal{P}}    =    \bar{p}_{0}   +    \frac{1}{\sqrt{8\mathcal{K}_{\mathcal{P}}}}\frac{\mathrm{d} W}{\mathrm{d} t},
\label{MomQuadHomoMeasurement}
\end{equation}
which assumes the value $\mathcal{P}$ with probability $P(\mathcal{P})$.  Substituting Eq.\ \eqref{MomQuadHomoMeasurement} in Eq.\ \eqref{CondAtmStateBeforeBS}, one can then obtain a continuous SSE with a Wiener noise.

Homodyne detection experiments can indeed be configured to access different quadratures of electromagnetic modes \cite{InOut1}. In Sec.\ \ref{Sec_HOMO_Measure}, however, we adopt an alternative approach by modeling the 50:50 beam splitter via the unitary matrix $U_\textrm{BS}$. This allows us to avoid introducing the continuum momentum basis $\{ \left|\mathcal{P}\right\rangle \}$ and eliminates the need for the undetermined parameter $\sigma$.

\subsection{Homodyne Measurement Signal}
\label{Sec_HOMO_Measure}

In this section, we provide a detailed treatment of the conventional homodyne measurement setup.  For this, we need to mix the output signal with a high intensity local oscillator.  We denote the (coarse-grained) creation operator for the strong local oscillator as $\hat{b}^{\dagger}$.  The coarse-grained modes at time $t$ of the local oscillator can be described by the coherent states $\left|\beta\right\rangle$, where $\hat{b}    \left|\beta\right\rangle  = \beta  \left|\beta\right\rangle$ and $\beta$ is real. 

We start by writing the collective density operator for the atomic state, the input mode, and the local oscillator mode at time $t$ as
\begin{equation}
\rho(t) = \left|\Psi(t)\right\rangle\left|\alpha\right\rangle\left|\beta\right\rangle\left\langle\beta\right|\left\langle\alpha\right|\left\langle\Psi(t)\right|.
\label{HomoDMTimet}
\end{equation}
Note that all three degrees of freedom are in a product state at this point. To write the conditioned density operator at time $t + \mathrm{d} t$, we first consider the unitary time evolution under the effective atom-input light Hamiltonian \eqref{EffHInputFin}.  We then describe the effect of the 50:50 beam splitter as the unitary operator $U_\textrm{BS}$. The effect of this operator is summarized as \cite{BeamSplitter}
\begin{equation}
U_\textrm{BS} \left|  \alpha_\mathrm{out}  \right\rangle \left|  \beta_\mathrm{LO}  \right\rangle  =  \left|\frac{  \alpha_\mathrm{out} + i\beta_\mathrm{LO}  }{\sqrt{2}}\right\rangle        \left|\frac{  \alpha_\mathrm{out} - i\beta_\mathrm{LO}  }{\sqrt{2}}\right\rangle,
\label{BeamSplitter}
\end{equation}
where $\left|  \alpha_\mathrm{out}  \right\rangle$ is the state corresponding to the output laser and $\left|  \beta_\mathrm{LO}  \right\rangle$ is the state of the local oscillator.  We recall that the only macroscopic information that is available to us is the difference between the photon numbers detected by the two detectors $\textrm{D}_1$ and $\textrm{D}_2$. This requires averaging over the possible detected photon numbers by $\textrm{D}_1$ and $\textrm{D}_2$ while keeping the difference constant.  

Denoting the photon number states detected by $\textrm{D}_1$ and $\textrm{D}_2$ as $\left|p+k\right\rangle$ and $\left|p\right\rangle$ respectively, we write the density operator at time $t+\mathrm{d} t$, conditioned on detecting $k$ photons, as
\begin{multline}
\rho(t+\mathrm{d} t) = \frac{1}{P_{k}}    \sum_{p}\left\langle p\right|    \left\langle p+k\right|    U_\textrm{BS}     e^{    -i\hat{\widetilde{\mathcal{H}}}_{\text{atm}}    \mathrm{d} t    }        e^{-i    \alpha_{0}\hat{\mathcal{M}} \left(  \hat{a}_{  \mathrm{in},  \mathrm{d} t  }    +    \hat{a}^{\dag}_{  \mathrm{in},  \mathrm{d} t  }     \right)    \mathrm{d} t    }     \\
\times    \left|\Psi(t)\right\rangle    \left|\alpha\right\rangle    \left|\beta\right\rangle    \left\langle\beta\right|    \left\langle\alpha\right|    \left\langle\Psi(t)\right|     e^{    +i    \alpha_{0}\hat{\mathcal{M}} \left(  \hat{a}_{  \mathrm{in},  \mathrm{d} t  }    +    \hat{a}^{\dag}_{  \mathrm{in},  \mathrm{d} t  }     \right)    \mathrm{d} t    }e^{    +i\hat{\widetilde{\mathcal{H}}}_{\text{atm}}       \mathrm{d} t    }
    U^{\dagger}_\textrm{BS}    \left|p+k\right\rangle    \left|p\right\rangle,
\label{HomoCondDM0}
\end{multline}
where $P_{k}$ is the probability of obtaining a measurement outcome $k$ that is derived from the normalization of $\rho(t+\mathrm{d} t)$.  

Following similar steps as in Eq.\ \eqref{StateBeforeBS},  we obtain
\begin{gather}
\begin{aligned}
& U_\textrm{BS}    e^{    -i\hat{\widetilde{\mathcal{H}}}_{\text{atm}}    \mathrm{d} t    }        e^{-i    \alpha_{0}\hat{\mathcal{M}} \left(  \hat{a}_{  \mathrm{in},  \mathrm{d} t  }    +    \hat{a}^{\dag}_{  \mathrm{in},  \mathrm{d} t  }     \right)    \mathrm{d} t    }    \left|\Psi\right\rangle\left|\alpha\right\rangle\left|\beta\right\rangle  \\
& = e^{    -i\hat{\widetilde{\mathcal{H}}}_{\text{atm}}    \mathrm{d} t    }     \sum_{m}C_{m}e^{-i\alpha_0\alpha m \mathrm{d} t}    \left|\psi_{m}\right\rangle      \left[    U_\textrm{BS}    \left|\alpha - i\alpha_0 m \mathrm{d} t\right\rangle    \left|\beta\right\rangle    \right] \\
& \overset{*}{=}     e^{    -i\hat{\widetilde{\mathcal{H}}}_{\text{atm}}    \mathrm{d} t    }     \sum_{m}C_{m}e^{-i\alpha_0\alpha m \mathrm{d} t}\left|\psi_{m}\right\rangle  \left|\alpha_{m}^{\prime}\right\rangle\left|\beta_{m}^{\prime}\right\rangle,
\end{aligned}
\label{StateAfterBS}
\end{gather}
where
\begin{equation}
\alpha_{m}^{\prime} = \frac{\alpha + i(\beta - \alpha_0 m \mathrm{d} t)}{\sqrt{2}}, \qquad \beta_{m}^{\prime} = \frac{\alpha - i(\beta + \alpha_0 m \mathrm{d} t)}{\sqrt{2}}.
\label{AlphaBetaPrime}
\end{equation}
To get to the last line in \eqref{StateAfterBS} (indicated by $*$), we have used Eq.\ \eqref{BeamSplitter}. Recall that $\beta^2 = \mathcal{O}(1/\mathrm{d} t) \gg 1$ is the number of photons contained in the single light mode of the local oscillator \cite{AEBN}. As a result, one also has $\left|\alpha_{m}^{\prime}\right|^2, \left|\beta_{m}^{\prime}\right|^2 \gg 1$. Using Eq.\ \eqref{StateAfterBS} into Eq.\ \eqref{HomoCondDM0}, one obtains
\begin{multline}
\rho(t+\mathrm{d} t) = \frac{1}{P_{k}}    \sum_{p,m,n}C_{m}C_{n}^{*}    \left\langle p\right|\left\langle p+k\right|    e^{    -i\hat{\widetilde{\mathcal{H}}}_{\text{atm}}    \mathrm{d} t    }     e^{    -i\alpha_0\alpha m \mathrm{d} t    }     \left|\psi_{m}\right\rangle    \left|\alpha_{m}^{\prime}\right\rangle    \left|\beta_{m}^{\prime}\right\rangle    \left\langle\beta_{n}^{\prime}\right|    \left\langle\alpha_{n}^{\prime}\right|    \left\langle\psi_{n}\right|     \\
\times     e^{    +i\hat{\widetilde{\mathcal{H}}}_{\text{atm}}    \mathrm{d} t    }    e^{    +i\alpha_0\alpha n \mathrm{d} t    }
\left|p+k\right\rangle\left|p\right\rangle.
\label{HomoCondDM1}
\end{multline}
Therefore, to calculate $P_{k}$, we trace over the atomic degrees of freedom in the numerator of Eq.\ \eqref{HomoCondDM1} and obtain
\begin{gather}
\label{PkInit1}
\begin{aligned}
P_{k}     &=     \textrm{Tr}_\text{atm}    \Bigg[    \sum_{p,m,n}C_{m}C_{n}^{*}    e^{-i\alpha_0\alpha (m - n) \mathrm{d} t}    \left\langle p\right|\left\langle p+k\right|    e^{    -i\hat{\widetilde{\mathcal{H}}}_{\text{atm}}    \mathrm{d} t    }    \left|\psi_{m}\right\rangle    \left|\alpha_{m}^{\prime}\right\rangle    \left|\beta_{m}^{\prime}\right\rangle       \\ 
&\qquad\qquad\qquad\qquad\qquad\qquad\qquad\qquad\qquad \times\left\langle\beta_{n}^{\prime}\right|    \left\langle\alpha_{n}^{\prime}\right|    \left\langle\psi_{n}\right|
 e^{    +i\hat{\widetilde{\mathcal{H}}}_{\text{atm}}    \mathrm{d} t    }    \left|p+k\right\rangle\left|p\right\rangle    \Bigg] \\
&=     \sum_{p,m,n}C_{m}C_{n}^{*}    e^{-i\alpha_0\alpha (m - n) \mathrm{d} t}    \left\langle p+k\right|\left.\alpha_{m}^{\prime}\right\rangle     \left\langle \alpha_{n}^{\prime} \right|\left. p+k\right\rangle     \left\langle p\right|\left.\beta_{m}^{\prime}\right\rangle     \left\langle \beta_{n}^{\prime} \right|\left. p\right\rangle   \\
&\qquad\qquad\qquad\qquad\qquad\qquad\qquad\qquad\qquad \times     \textrm{  Tr}_\text{atm}    \left[    e^{    -i\hat{\widetilde{\mathcal{H}}}_{\text{atm}}    \mathrm{d} t    }    \left|\psi_{m}\right\rangle\left\langle\psi_{n}\right|     e^{    +i\hat{\widetilde{\mathcal{H}}}_{\text{atm}}    \mathrm{d} t    }      \right]   \\
&= \sum_{p,m}\left|C_{m}\right|^{2}\left|\left\langle p+k\right|\left.\alpha_{m}^{\prime}\right\rangle\right|^{2} \left| \left\langle p\right|\left.\beta_{m}^{\prime}\right\rangle \right|^{2} \\
&= \sum_{m}\left|C_{m}\right|^{2}\sum_{p = 0}^{\infty}\frac{1}{2\pi\left|\alpha_{m}^{\prime}\right|\left|\beta_{m}^{\prime}\right|}\exp{\left(-\frac{(p + k - \left|\alpha_{m}^{\prime}\right|^2)^{2}}{2\left|\alpha_{m}^{\prime}\right|^{2}} - \frac{(p-\left|\beta_{m}^{\prime}\right|^{2})^{2}}{2\left|\beta_{m}^{\prime}\right|^{2}}\right)},
\end{aligned}
\end{gather}
where in the limit $\mu\gg 1$ we have used the approximation
\begin{equation}
\frac{1}{n!}\mu^{n}e^{-\mu} \approx \frac{1}{\sqrt{2\pi\mu}}e^{-(n-\mu)^{2}/2\mu}.
\label{PoissonGaussian}
\end{equation}
We compute the summation over $p$ in the last line of Eq.\ \eqref{PkInit1} in \ref{AppnGaussianPk}.  The steps are as follows:
\begin{enumerate}
\item Complete the square for the quadratic polynomial in $p$ inside the argument of the exponential.
\item Replace the summation $\sum_{p=0}^{\infty}$ by the integral $\int_{0}^{\infty}\mathrm{d} p$.
\item After introducing a variable change, perform the Gaussian integral.
\end{enumerate}  

Using  Eq.\ \eqref{SimplifyPk3Gaussian122} in Eq.\ \eqref{PkInit1}, we now find
\begin{gather}
\label{PkInit2}
\begin{aligned}
P_{k} &\approx \sum_{m}\left|C_{m}\right|^{2}\frac{1}{\sqrt{2\pi\left(\left|\beta_{m}^{\prime}\right|^{2} + \left|\alpha_{m}^{\prime}\right|^{2}\right)}}\exp{\left[-\frac{\left(k + \left|\beta_{m}^{\prime}\right|^{2} - \left|\alpha_{m}^{\prime}\right|^{2}\right)^{2}}{2\left(\left|\beta_{m}^{\prime}\right|^{2} + \left|\alpha_{m}^{\prime}\right|^{2}\right)}\right]} \\
&= \sum_{m}\left|C_{m}\right|^{2}\frac{1}{\sqrt{2\pi\left[ \alpha^{2} + \beta^{2} +  \alpha^{2}_{0}m^2\mathrm{d} t^2 \right]}}\exp{\left(-\frac{\left[k + 2\alpha_0\beta m \mathrm{d} t\right]^{2}}{2\left[ \alpha^{2} + \beta^{2} +  \alpha^{2}_{0}m^2\mathrm{d} t^2 \right]}\right)}.
\end{aligned}
\end{gather}
Neglecting terms $\mathcal{O}(\mathrm{d} t^2)$ in Eq.\ \eqref{PkInit2}, we write
\begin{equation}
\label{PkInit3}
\begin{split}
P_{k} &\approx \sum_{m}\left|C_{m}\right|^{2}\frac{1}{\sqrt{2\pi\left[ \alpha^{2} + \beta^{2}\right]}}    \exp{\left(-\frac{\left[k + 2\alpha_0\beta m \mathrm{d} t\right]^{2}}{2\left[ \alpha^{2} + \beta^{2}\right]}\right)} \\
& \overset{*1}{\approx}     \frac{1}{\sqrt{2\pi\left[ \alpha^{2} + \beta^{2}\right]}}    \exp{\left(-\frac{   k^2 + 4k\alpha_0\beta m \mathrm{d} t   }{2\left[ \alpha^{2} + \beta^{2}\right]}\right)} \\
& \overset{*2}{\approx}     \sum_{m}\left|C_{m}\right|^{2}\frac{1}{\sqrt{2\pi\left[ \alpha^{2} + \beta^{2}\right]}}   e^{-\frac{k^2}{2\left[ \alpha^{2} + \beta^{2}\right]}}   \left(1 -\frac{4k\alpha_0\beta m \mathrm{d} t}{2\left[ \alpha^{2} + \beta^{2}\right]}\right)\\
&=     \sum_{m}\left|C_{m}\right|^{2}\frac{1}{\sqrt{2\pi\left[ \alpha^{2} + \beta^{2}\right]}}e^{-\frac{k^2}{2\left[ \alpha^{2} + \beta^{2}\right]}}\left(1 -\frac{4k\alpha_0\beta \left\langle \hat{\mathcal{M}} \right\rangle \mathrm{d} t}{2\left[ \alpha^{2} + \beta^{2}\right]}\right)\\
&\approx \frac{1}{\sqrt{2\pi\left[ \alpha^{2} + \beta^{2}\right]}}\exp{\left(-\frac{\left[k + 2\alpha_0\beta \left\langle \hat{\mathcal{M}} \right\rangle \mathrm{d} t\right]^{2}}{2\left[ \alpha^{2} + \beta^{2}\right]}\right)},
\end{split}
\end{equation}
where, in the second line ($*1$) of Eq.\ \eqref{PkInit3}, we kept terms upto $\mathcal{O}(\mathrm{d} t)$.  Since $\beta = \mathcal{O}(1/\sqrt{\mathrm{d} t} ) \gg 1,$ we kept terms upto $\mathcal{O}\left(\alpha_0\beta/\left( \alpha^{2} + \beta^{2}\right)\right)$ in the third line ($*2$) of Eq.\ \eqref{PkInit3}.  Another justification of the same result is that in the first line of Eq.\ \eqref{PkInit3} the variance of the Gaussian $2\left( \alpha^{2} + \beta^{2}\right)$ is much broader than the functional dependence of $\left|C_{m}\right|^{2}$ on $m$. This allows us to use \cite{JS}
\begin{equation}
\label{CmDelta}
\left| C_{m}\right|^{2} \approx \delta_{m, \left\langle \hat{\mathcal{M}} \right\rangle}.
\end{equation}
Note that the steps in Eq.\ \eqref{PkInit3} are similar to Eq.\ \eqref{MomQuadProb}.  However, in Eq.\ \eqref{PkInit3}, we have to consider taking two limits: $\mathrm{d} t \rightarrow 0$ and $\beta \rightarrow \infty$.    

Similar to Eq.\ \eqref{MomQuadWienerDefn}, we identify from Eq.\ \eqref{PkInit3} 
\begin{equation}
\mathrm{d} W = \sqrt{\mathrm{d} t}\frac{\left(k + 2\alpha_0\beta \left\langle \hat{\mathcal{M}} \right\rangle \mathrm{d} t\right)}{\sqrt{\alpha^{2} + \beta^{2}}},
\label{WienerDefn}
\end{equation}
where $\mathrm{d} W$ is the Wiener process. Using this, we obtain the stochastic variable
\begin{equation}
\hat{k} = \left[ \sqrt{\mathrm{d} t\left (\alpha^{2} + \beta^{2}\right)} \right]\frac{\mathrm{d} W}{\mathrm{d} t} - 2\alpha_0\beta \left\langle \hat{\mathcal{M}} \right\rangle \mathrm{d} t,
\label{HomoMeasurement}
\end{equation}
that assumes the value $k$ with probability $P_{k}$.   Defining
\begin{equation}
\label{RescaleHomoDeriv}
\mathcal{K} = \frac{1}{8\mathrm{d} t\left(\alpha^{2} + \beta^{2}\right)}, \qquad \hat{\mathcal{M}}_\textrm{r} = -2\alpha_{0}\beta \mathrm{d} t \hat{\mathcal{M}},
\end{equation}
we write Eq.\ \eqref{HomoMeasurement} as
\begin{equation}
\hat{k} = \left\langle \hat{\mathcal{M}}_\textrm{r} \right\rangle + \frac{1}{\sqrt{8\mathcal{K}}}\frac{\mathrm{d} W}{\mathrm{d} t},
\label{Hom_Curr_JS}
\end{equation}
which is the same as Eq.\ (26) of Ref.\ \cite{JS}.  This variable $\hat{k}$ is later related to the homodyne measurement signal --- see Eqs.\ \eqref{SSEPRA_Signal} and \eqref{FinalPres}.

\subsection{Stochastic Schr\"{o}dinger Equation for Homodyne Measurement Setup}
\label{Sec_SSE}

We want to write the density matrix $\rho(t+\mathrm{d} t)$ as a rank-1 projector, which can then be interpreted as the outer product of the state $\left|\Psi(t+\mathrm{d} t)\right\rangle$ with itself.   We first simplify the inner products like $\left\langle p\right|\left.\beta_{m}^{\prime}\right\rangle$ using the fact $\left|\alpha_{m}^{\prime}\right|^2,  \left|\alpha_{n}^{\prime}\right|^2, \left|\beta_{m}^{\prime}\right|^2, \left|\beta_{n}^{\prime}\right|^2 \gg 1$.  To that end,  we take the square root of Eq.\ \eqref{PoissonGaussian} and obtain
\begin{equation}
\left\langle p\right|\left.\beta_{m}^{\prime}\right\rangle  = \frac{\left(\beta_{m}^{\prime}\right)^{p}}{\sqrt{p!}}e^{ -\left|\beta_{m}^{\prime}\right|^2 / 2   }    
\approx \frac{1}{\left[2\pi\left|\beta_{m}^{\prime}\right|^2\right]^{\frac{1}{4}}}e^{-\frac{(p-\left|\beta_{m}^{\prime}\right|^2)^{2}}{4\left|\beta_{m}^{\prime}\right|^2}}e^{ip\mathrm{Arg} (\beta_{m}^{\prime})}.
\label{PoissonGaussian1/2}
\end{equation}
We now simplify Eq.\ \eqref{HomoCondDM1} using  Eq.\ \eqref{PoissonGaussian1/2} as
\begin{gather}
\label{HomoCondDM2}
\begin{aligned}
\rho(t+\mathrm{d} t) &= \left|\Psi(t+\mathrm{d} t)\right\rangle        \left\langle\Psi(t+\mathrm{d} t)\right| \\
&=  \frac{1}{P_{k}}          \sum_{m,n}\left[    C_{m}C_{n}^{*}        e^{        -i\hat{\widetilde{\mathcal{H}}}_{\text{atm}}       \mathrm{d} t        }          e^{        -i\alpha_0\alpha m \mathrm{d} t        }  \left|\psi_{m}\right\rangle         \left\langle\psi_{n}\right|        e^{        +i\hat{\widetilde{\mathcal{H}}}_{\text{atm}}           \mathrm{d} t        }  e^{       +i\alpha_0\alpha n \mathrm{d} t       }    \right]   \\ 
&\qquad\qquad\qquad\qquad\qquad\qquad\qquad
\times\sum_{p}    \left\langle p\right|\left.\beta_{m}^{\prime}\right\rangle    \left\langle \beta_{n}^{\prime}\right|\left. p \right\rangle    \left\langle p+k\right|\left.\alpha_{m}^{\prime}\right\rangle    \left\langle \alpha_{n}^{\prime}\right|\left. p+k \right\rangle   \\
&\approx        \sum_{m,n}    \frac{        C_{m}C_{n}^{*}    e^{-i\alpha_0\alpha (m-n) \mathrm{d} t}        }{        2\pi P_{k}\left[    \left|\beta_{m}^{\prime}\right|^2\left|\beta_{n}^{\prime}\right|^2 \left|\alpha_{m}^{\prime}\right|^2\left|\alpha_{n}^{\prime}\right|^2    \right]    ^{\frac{1}{4}}       } \left[        e^{        -i    \hat{\widetilde{\mathcal{H}}}_{\text{atm}}          \mathrm{d} t        }        \left|\psi_{m}\right\rangle \left\langle\psi_{n}\right|        e^{        +i        \hat{\widetilde{\mathcal{H}}}_{\text{atm}}          \mathrm{d} t        }       \right] \\
   &\qquad \times    \sum_{p}     \exp  \left[    ip\bigg(   \mathrm{Arg} (\beta_{m}^{\prime}) + \mathrm{Arg} (\beta_{n}^{\prime *}) + \mathrm{Arg} (\alpha_{m}^{\prime}) + \mathrm{Arg} (\alpha_{n}^{\prime *})    \bigg)    \right]  \\
  &\qquad\qquad \times   \exp \left[    -\frac{(p-\left|\beta_{m}^{\prime}\right|^2)^{2}}{4\left|\beta_{m}^{\prime}\right|^2}    -\frac{(p-\left|\beta_{n}^{\prime}\right|^2)^{2}}{4\left|\beta_{n}^{\prime}\right|^2}    -\frac{(p + k -\left|\alpha_{m}^{\prime}\right|^2)^{2}}{4\left|\alpha_{m}^{\prime}\right|^2}    -\frac{(p + k -\left|\alpha_{n}^{\prime}\right|^2)^{2}}{4\left|\alpha_{n}^{\prime}\right|^2}    \right] \\
&\approx \sum_{m,n}    \frac{    C_{m}C_{n}^{*}e^{    -i\alpha_0\alpha (m-n) \mathrm{d} t    }      \left[        e^{        -i    \hat{\widetilde{\mathcal{H}}}_{\text{atm}}          \mathrm{d} t        }        \left|\psi_{m}\right\rangle \left\langle\psi_{n}\right|        e^{        +i        \hat{\widetilde{\mathcal{H}}}_{\text{atm}}          \mathrm{d} t        }       \right]    }{2\pi P_{k}\left[\left|\beta_{m}^{\prime}\right|^2\left|\beta_{n}^{\prime}\right|^2 \left|\alpha_{m}^{\prime}\right|^2\left|\alpha_{n}^{\prime}\right|^2\right]^{\frac{1}{4}}}      \\
&\qquad\qquad\qquad \times \sum_{p}    \exp \left[    -\frac{(p-\left|\beta_{m}^{\prime}\right|^2)^{2}}{4\left|\beta_{m}^{\prime}\right|^2}    -\frac{(p-\left|\beta_{n}^{\prime}\right|^2)^{2}}{4\left|\beta_{n}^{\prime}\right|^2} \right. \\
& \left.   \qquad\qquad\qquad\qquad\qquad\qquad -\frac{(p + k -\left|\alpha_{m}^{\prime}\right|^2)^{2}}{4\left|\alpha_{m}^{\prime}\right|^2}    -\frac{(p + k -\left|\alpha_{n}^{\prime}\right|^2)^{2}}{4\left|\alpha_{n}^{\prime}\right|^2}    - p\theta    \right],
\end{aligned}
\end{gather}
where, assuming $-\pi/2  \leqslant  \mathrm{Arg} (\beta_{m}^{\prime}), \mathrm{Arg} (\beta_{n}^{\prime *}), \mathrm{Arg} (\alpha_{m}^{\prime}), \mathrm{Arg} (\alpha_{n}^{\prime *}) <+\pi/2$, we have defined
\begin{equation}
\label{ExpArg}
i\theta = \mathrm{Arg} (\beta_{m}^{\prime}) + \mathrm{Arg} (\beta_{n}^{\prime *})  + \mathrm{Arg} (\alpha_{m}^{\prime}) + \mathrm{Arg} (\alpha_{n}^{\prime *}).
\end{equation}
In \ref{ValueOfITheta}, we show that upto $\mathcal{O}(\mathrm{d} t)$ and $\mathcal{O}\left(\alpha_0\beta/\left( \alpha^{2} + \beta^{2}\right)\right)$
\begin{equation}
\label{4ExpArg}
\theta \approx -2i\tan^{-1}\left( \frac{\alpha\alpha_0(n-m)\mathrm{d} t}{ \alpha^{2} + \beta^{2}  }   \right).
\end{equation}
In \ref{AppnGaussianrhotdt},  we calculate the summation over $p$ that appears in Eq.\ \eqref{HomoCondDM2}, see Eq.\ \eqref{SimplifyHomoCondDMGaussian3}.

We use the result of Eq.\ \eqref{SimplifyHomoCondDMGaussian3} in Eq.\ \eqref{HomoCondDM2}.  In \ref{AppnDM}, we simplify the density matrix $\rho(t+\mathrm{d} t)$ further by neglecting $\mathcal{O}(\mathrm{d} t^2)$ terms and using $\beta^2 = \mathcal{O}(1/\mathrm{d} t) \gg 1$.   As a result,  we derive
\begin{multline}
\label{HomoCondDMFinal}
\rho(t+\mathrm{d} t)     =     \left|\Psi(t+\mathrm{d} t)\right\rangle\left\langle\Psi(t+\mathrm{d} t)\right|        \\
\propto     e^{    -i    \hat{\mathcal{H}}_\text{atm}^\text{eff}    \mathrm{d} t    }      e^{    -\frac{\left[\hat{k} + 2\alpha_0\beta \hat{\mathcal{M}} \mathrm{d} t\right]^{2}}{4\left[ \alpha^{2} + \beta^{2}\right]}    }  \left|\Psi(t)\right\rangle        \left\langle\Psi(t)\right|      e^{        -\frac{\left[\hat{k} + 2\alpha_0\beta \hat{\mathcal{M}} \mathrm{d} t\right]^{2}}{4\left[ \alpha^{2} + \beta^{2}\right]}        }          e^{        +i        \hat{\mathcal{H}}_\text{atm}^\text{eff} \mathrm{d} t        },
\end{multline} 
where 
\begin{gather}
\begin{aligned}
\hat{\mathcal{H}}_\text{atm}^\text{eff}     &=     \hat{\widetilde{\mathcal{H}}}_{\text{atm}} + 2 \alpha_0\alpha \hat{\mathcal{M}} \\
&=     \hat{\mathcal{H}}_{\text{atm}}      - \frac{\alpha}{\beta \mathrm{d} t} \hat{\mathcal{M}}_\textrm{r}      +       \frac{   \left| \mathcal{A}_0 \right|^{2} \omega_{L}  }{  4\alpha_0^2 \beta^2  \kappa^{2} \mathrm{d} t^2       }  \hat{\mathcal{M}}_\textrm{r}^{2}.
\end{aligned}
\label{UtmostFinalEffHMainText}
\end{gather}
Equation \eqref{HomoCondDMFinal} obtains the following expression for the changed quantum state after a single weak measurement in the time step $\mathrm{d} t$:
\begin{equation}
\label{HomoCondWF}
\left|\Psi(t+\mathrm{d} t)\right\rangle    \propto    e^{    -i    \hat{\mathcal{H}}_\text{atm}^\text{eff}    \mathrm{d} t    }          e^{        -\frac{\left[\hat{k} + 2\alpha_0\beta \hat{\mathcal{M}} \mathrm{d} t\right]^{2}}{4\left[ \alpha^{2} + \beta^{2}\right]}        }          \left|\Psi(t)\right\rangle         =         e^{    -i    \hat{\mathcal{H}}_\text{atm}^\text{eff}    \mathrm{d} t    }          e^{        - 2\mathcal{K}\mathrm{d} t \left(\hat{k} - \hat{\mathcal{M}}_\textrm{r} \right)^{2}        }  \left|\Psi(t)\right\rangle,
\end{equation}
where we used the definitions of Eq.\ \eqref{RescaleHomoDeriv}. This expression is the same as Eq.\ (27) of Ref.\ \cite{JS}. Here we also delineate the generalized Gaussian measurement operators and the corresponding positive operator-valued measures.  Following the same procedure as in Ref.\ \cite{JS}, we obtain the SSE
\begin{equation}
\mathrm{d} \left | \bar{\Psi}\left(t\right) \right \rangle  = \left\lbrace     - i        \hat{\mathcal{H}}_\text{atm}^\text{eff}   \mathrm{d} t         -\mathcal{K} \hat{\mathcal{M}}_\textrm{r}^{2}\mathrm{d} t      + 4\mathcal{K}\hat{\mathcal{M}}_\textrm{r}    \hat{k}\mathrm{d} t    \right\rbrace \left | \bar{\Psi}\left(t\right) \right \rangle,
\label{SSE_Blatt}
\end{equation}
where $\hat{k}$ was defined in Eq.\ \eqref{Hom_Curr_JS}.  In general, the Hamiltonian $\hat{\mathcal{H}}_\text{atm}^\text{eff}$ contains perturbative correction terms that are linear and quadratic in the measurement operator $\hat{\mathcal{M}}_\textrm{r}$.    To obtain the measurement signal and the SSE considered in Ref.\ \cite{APLBJJ}
\begin{subequations}
\begin{align}
I(t)      &=      2\gamma \langle \hat{\mathcal{M}_0} \rangle + \sqrt{\gamma} \: \frac{\mathrm{d} W}{\mathrm{d} t},   \label{SSEPRA_Signal}  \\
\text{d}|\bar{\Psi}(t)\rangle      &=      \Bigl[-i\hat{\mathcal{H}}_\text{atm} - \frac{\gamma}{2} \hat{\mathcal{M}}_0^2 + I(t) \hat{\mathcal{M}}_0 \Bigr]\text{d}t|\bar{\Psi}(t)\rangle,
\end{align}
\label{SSEPRA}
\end{subequations}
one needs to operate in the limit of $\alpha\ll 1$ and $\kappa \gg 1$. Recall that $2\kappa$ is the cavity mode decay rate, where as the constant $\mathcal{K} = 1/[8\mathrm{d} t\left(\alpha^{2} + \beta^{2}\right)]$ is proportional to the ``measurement strength" \cite{Sherson_PRA, APLBJJ}. In this limit, we have $\hat{\mathcal{H}}_\text{atm}^\text{eff} \approx \hat{\mathcal{H}}_\text{atm}$. We have further used the following redefinition:
\begin{equation}
\label{FinalPres}
\textrm{define: } I(t) = 16\mathcal{K}\hat{k} = \frac{\gamma}{2}\hat{k}, \qquad \textrm{replace: }\hat{\mathcal{M}}_\textrm{r} \rightarrow 4\hat{\mathcal{M}}_{0}, 32\mathcal{K} \rightarrow \gamma.
\end{equation}

\begin{figure}
\begin{indented}\item[]
\includegraphics[trim={0.5cm 7cm 0.5cm 7cm},clip, width=0.45\textwidth]{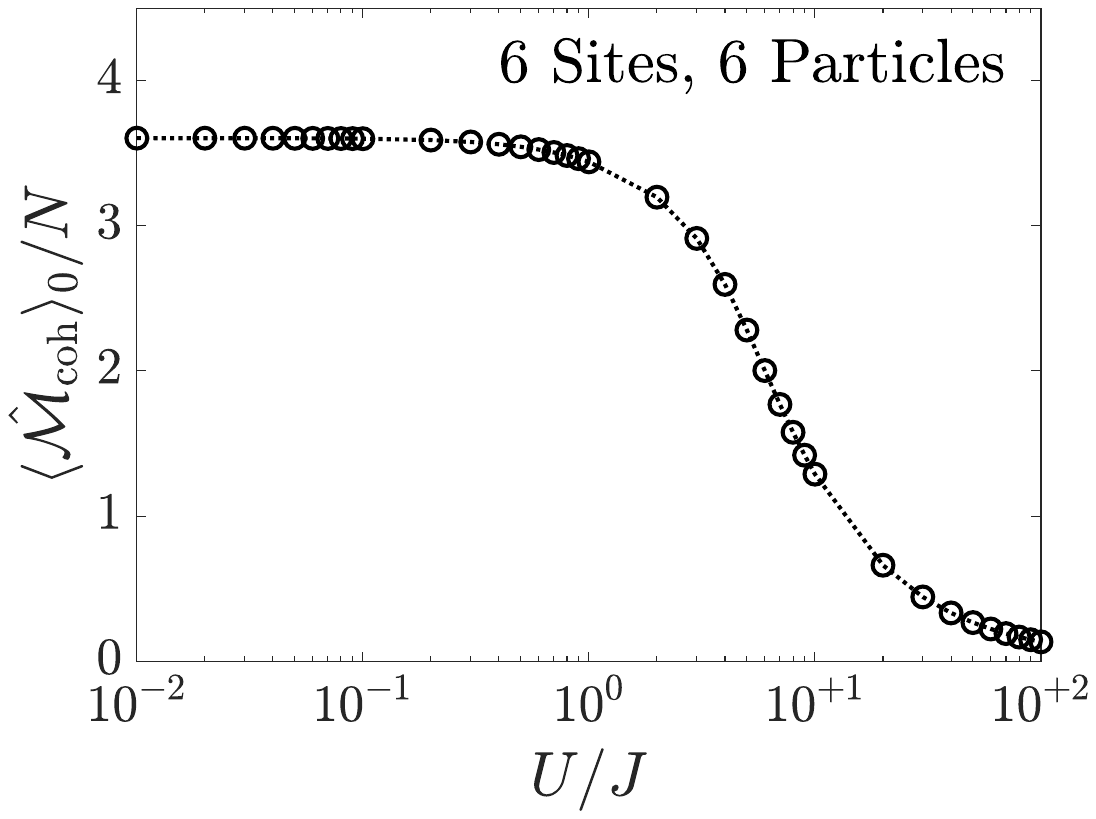}
\caption{The variation of $\langle \hat{\mathcal{M}}_\mathrm{coh} \rangle_0/N$ with $U/J$ for a Bose-Hubbard system with six sites and six particles.  Here we have $\langle \hat{\mathcal{M}}_\mathrm{coh} \rangle_0 = |\langle \psi_0 | \hat{\mathcal{M}}_\mathrm{coh} | \psi_0 \rangle|$ with $| \psi_0 \rangle$ being the ground state of the Bose-Hubbard Hamiltonian \eqref{BH_H}. This plot, similar to the plot of the condensate fraction in Ref.\ \cite{APLBJJ}, indicates a superfluid to Mott-insulator transition for $U/J$ between $1$ and $10$.}
\label{FigCohOrderParameter}
\end{indented}
\end{figure}

\begin{figure}
\includegraphics[trim={0.5cm 7cm 0.5cm 7cm},clip, width=0.33\textwidth]{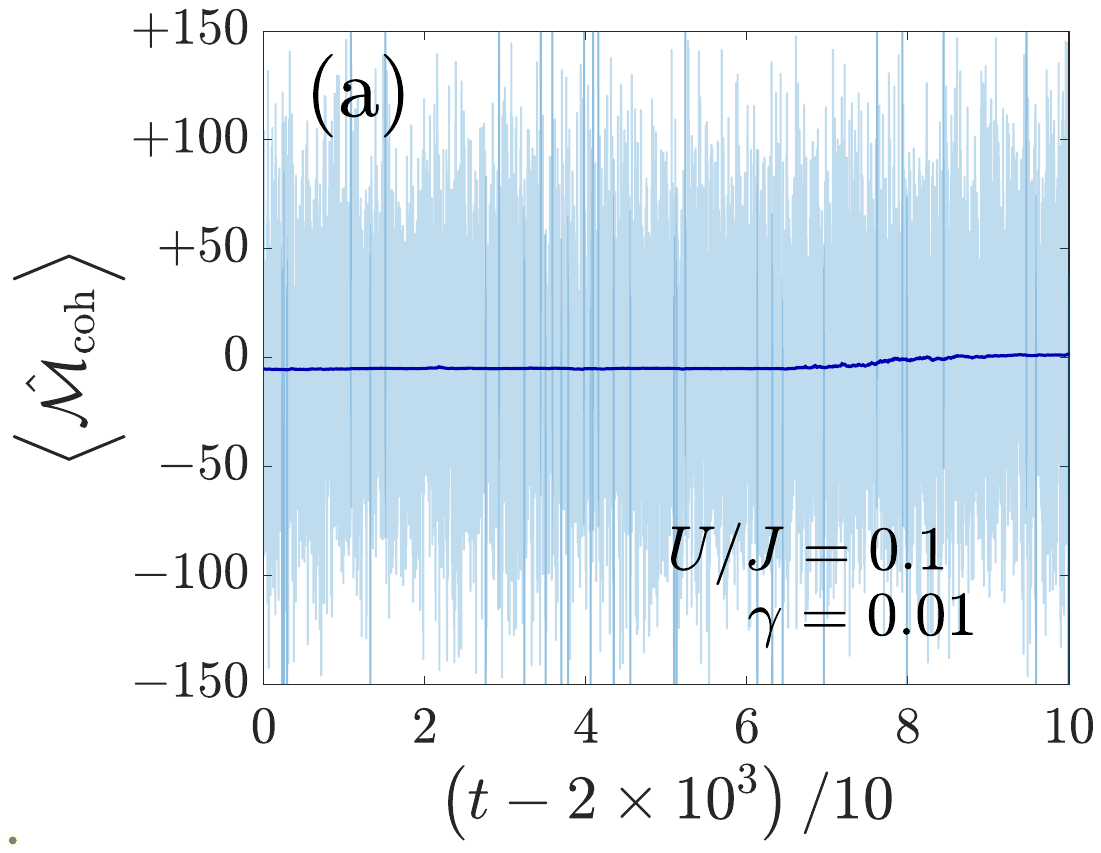}\hfill
\includegraphics[trim={0.5cm 7cm 0.5cm 7cm},clip, width=0.33\textwidth]{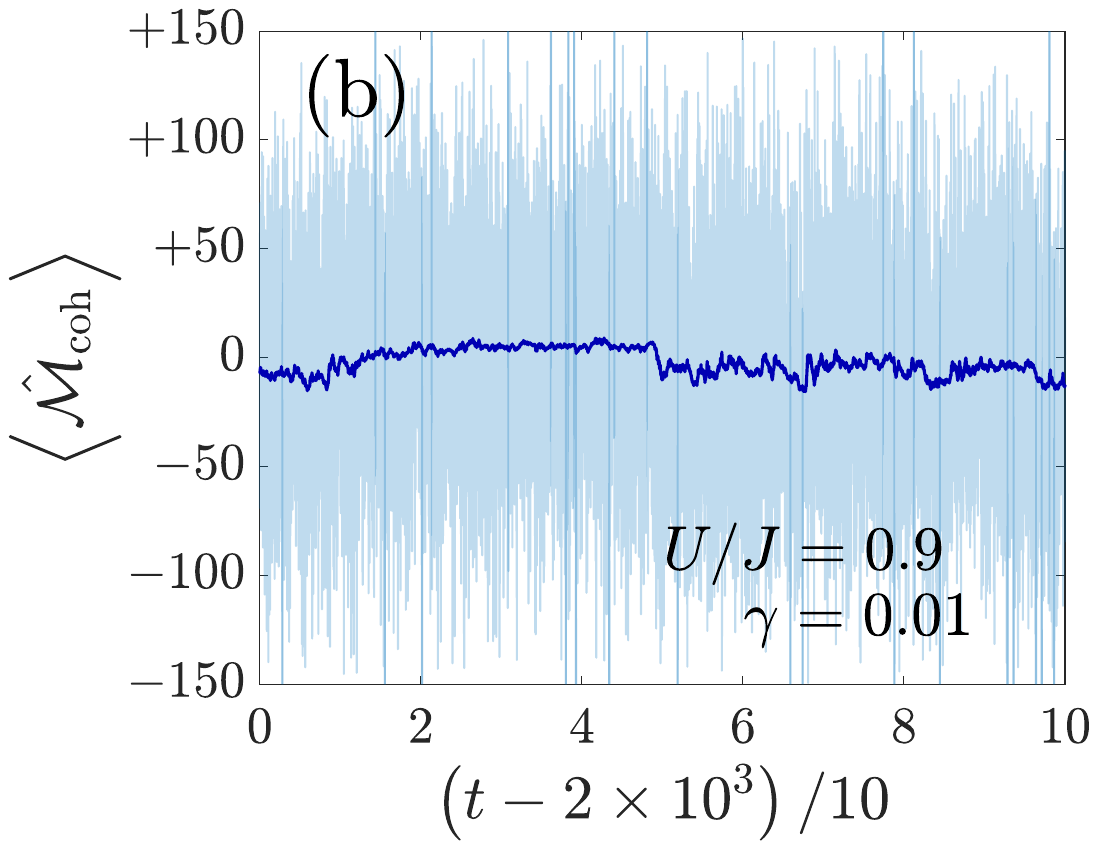}\hfill
\includegraphics[trim={0.5cm 7cm 0.5cm 7cm},clip, width=0.33\textwidth]{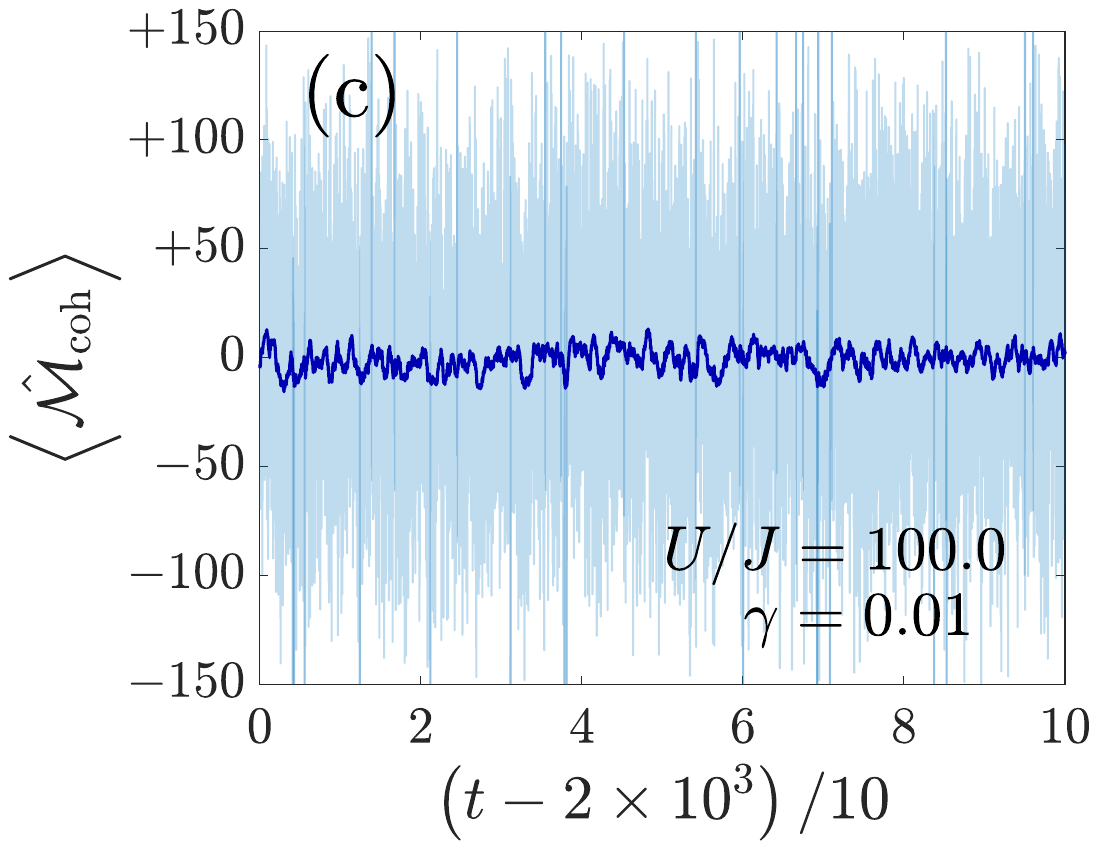}      \\
\includegraphics[trim={0.5cm 7cm 0.5cm 7cm},clip, width=0.33\textwidth]{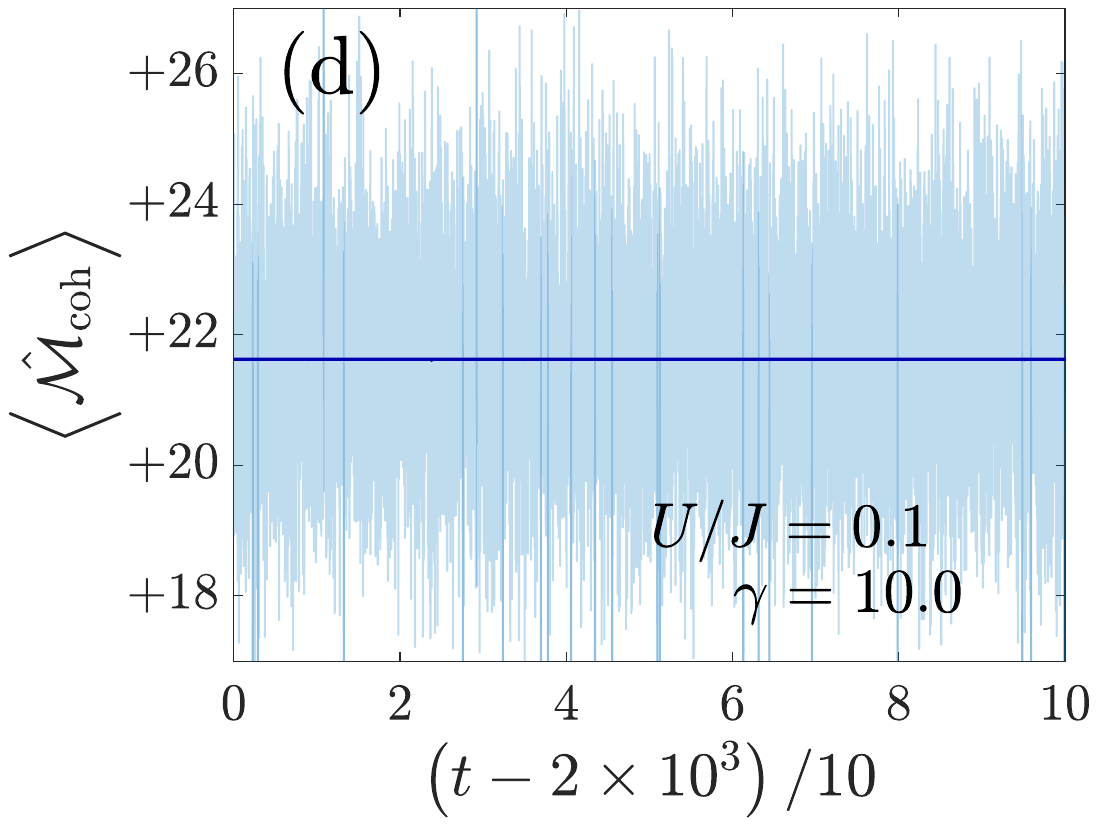}\hfill
\includegraphics[trim={0.5cm 7cm 0.5cm 7cm},clip, width=0.33\textwidth]{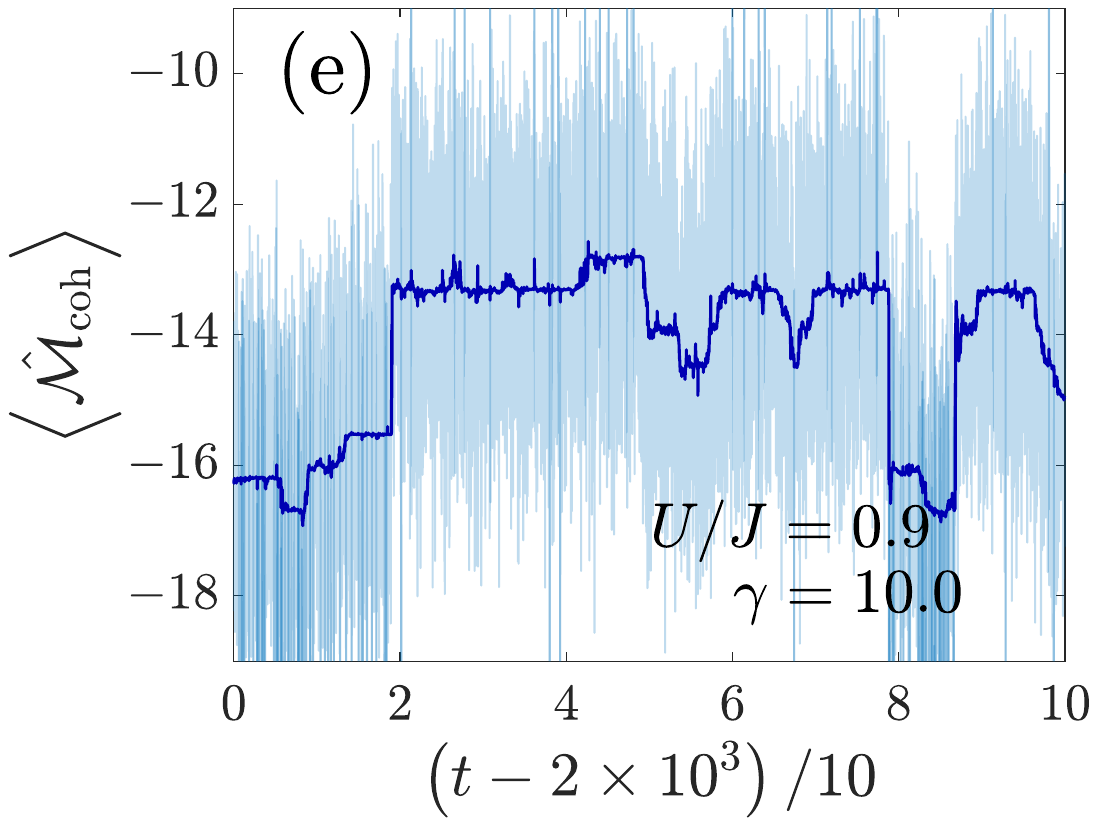}\hfill
\includegraphics[trim={0.5cm 7cm 0.5cm 7cm},clip, width=0.33\textwidth]{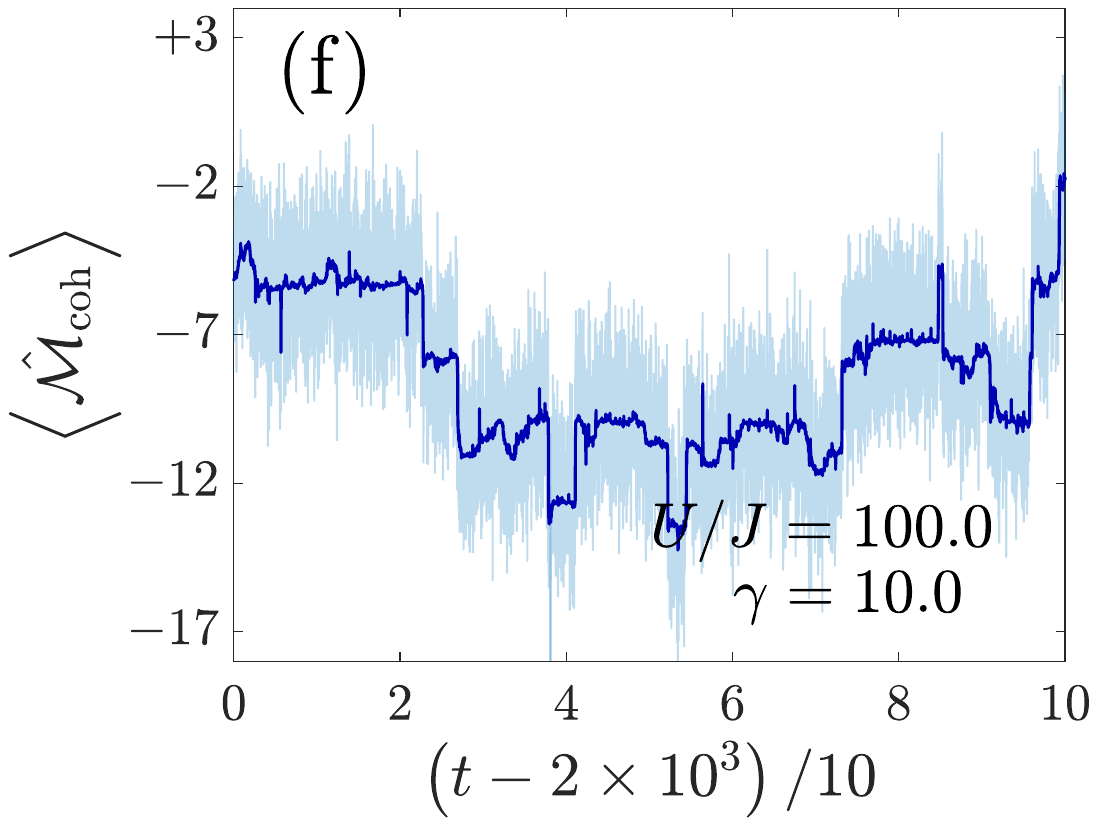}
\caption{We plot the time dependent expectation value \eqref{ExptM0PlotDefn} of the measurement operator $\hat{\mathcal{M}}_\mathrm{coh}$ vs time in dark blue. We also plot $I(t)/(2\gamma)$ in faded blue to show the effect of the Wiener noise in the homodyne measurement signal. We see the signature of the superfluid to Mott-insulator phase transition in the time dependent dynamics of $\left\langle \hat{\mathcal{M}}_\mathrm{coh} \right\rangle$ in panels \textbf{(a)}, \textbf{(b)} and \textbf{(c)}. Considering $10^3$ times larger measurement strength in panels \textbf{(d)}, \textbf{(e)} and \textbf{(f)}, we show examples of quantum trajectories in the quantum Zeno regime. Deep in the Mott-insulating phase, we observe frequent quantum jumps in this regime.}
\label{FigCohDynamics}
\end{figure}

\begin{figure}
\includegraphics[trim={0.5cm 7cm 0.5cm 7cm},clip, width=0.33\textwidth]{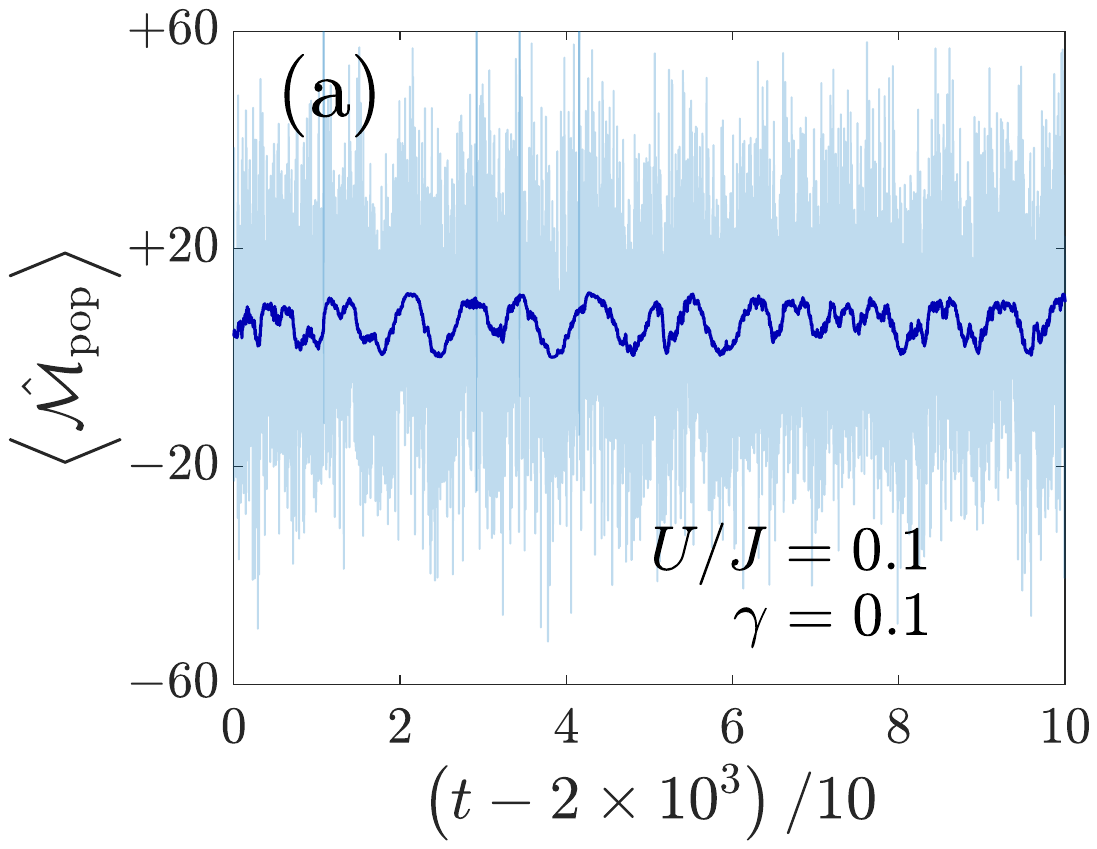}\hfill
\includegraphics[trim={0.5cm 7cm 0.5cm 7cm},clip, width=0.33\textwidth]{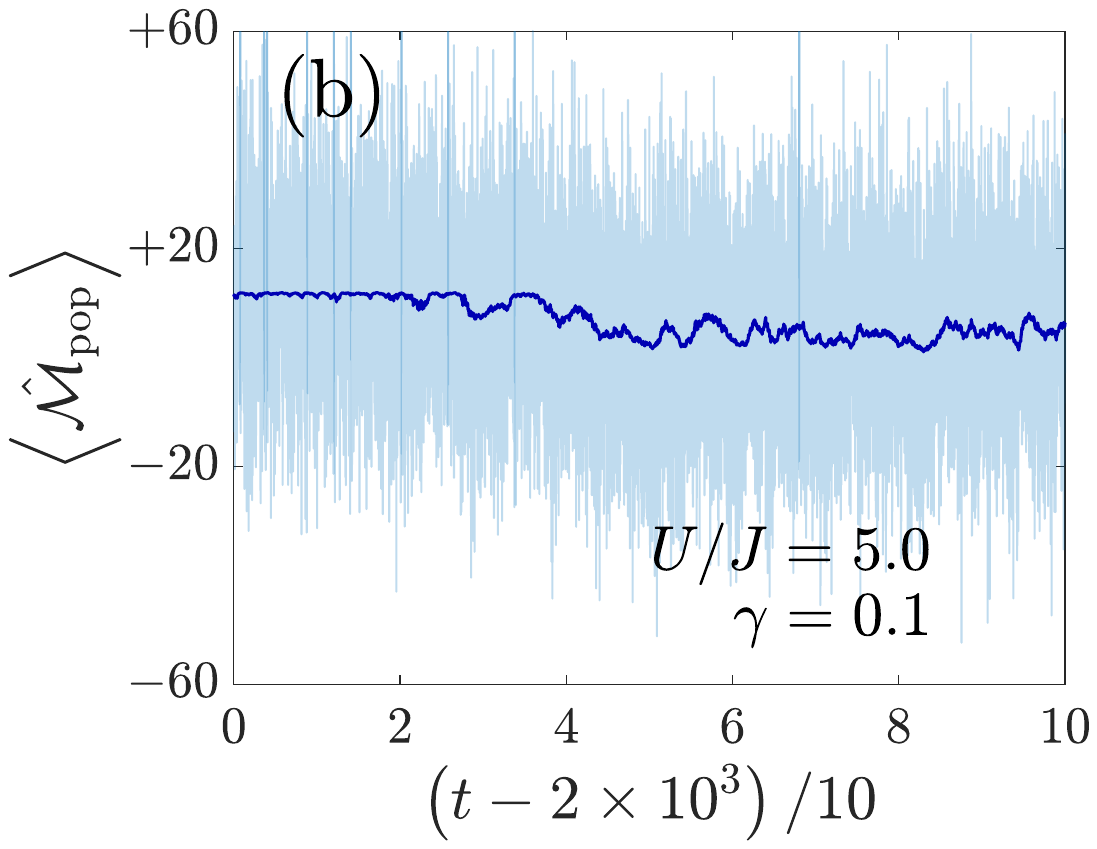}\hfill
\includegraphics[trim={0.5cm 7cm 0.5cm 7cm},clip, width=0.33\textwidth]{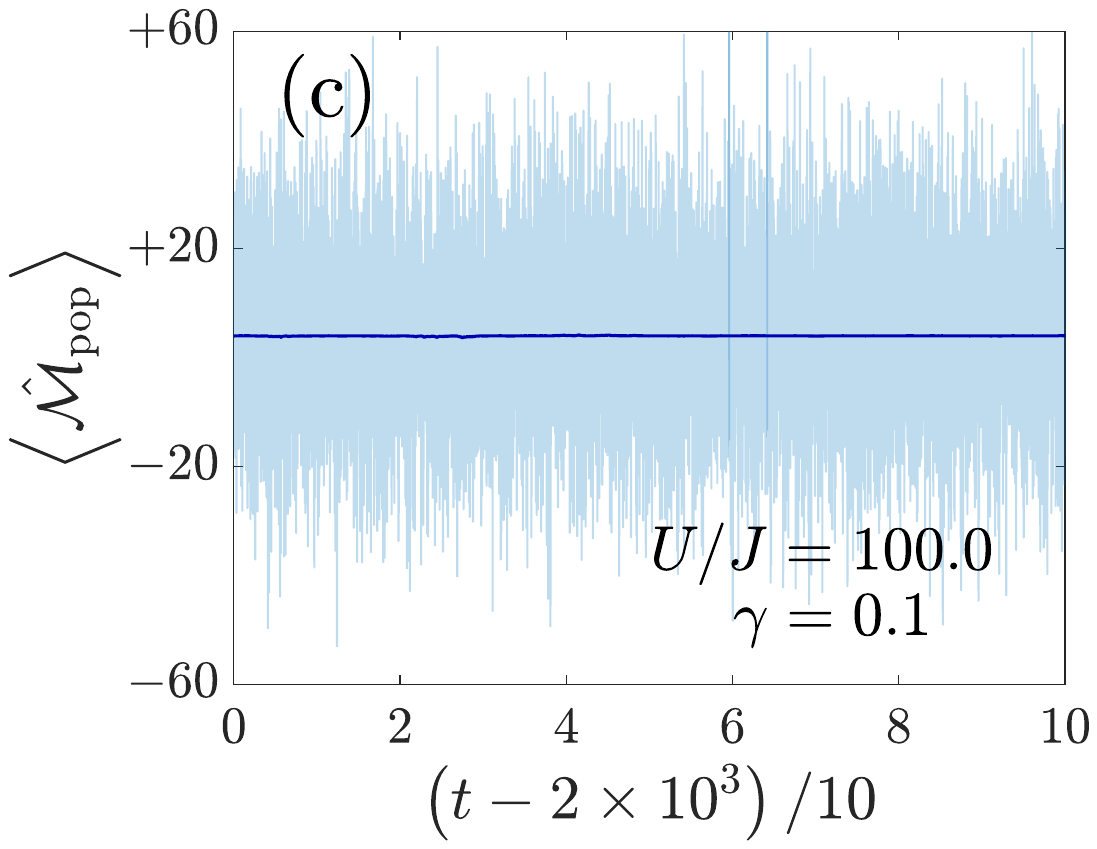} \\
\includegraphics[trim={0.5cm 7cm 0.5cm 7cm},clip, width=0.33\textwidth]{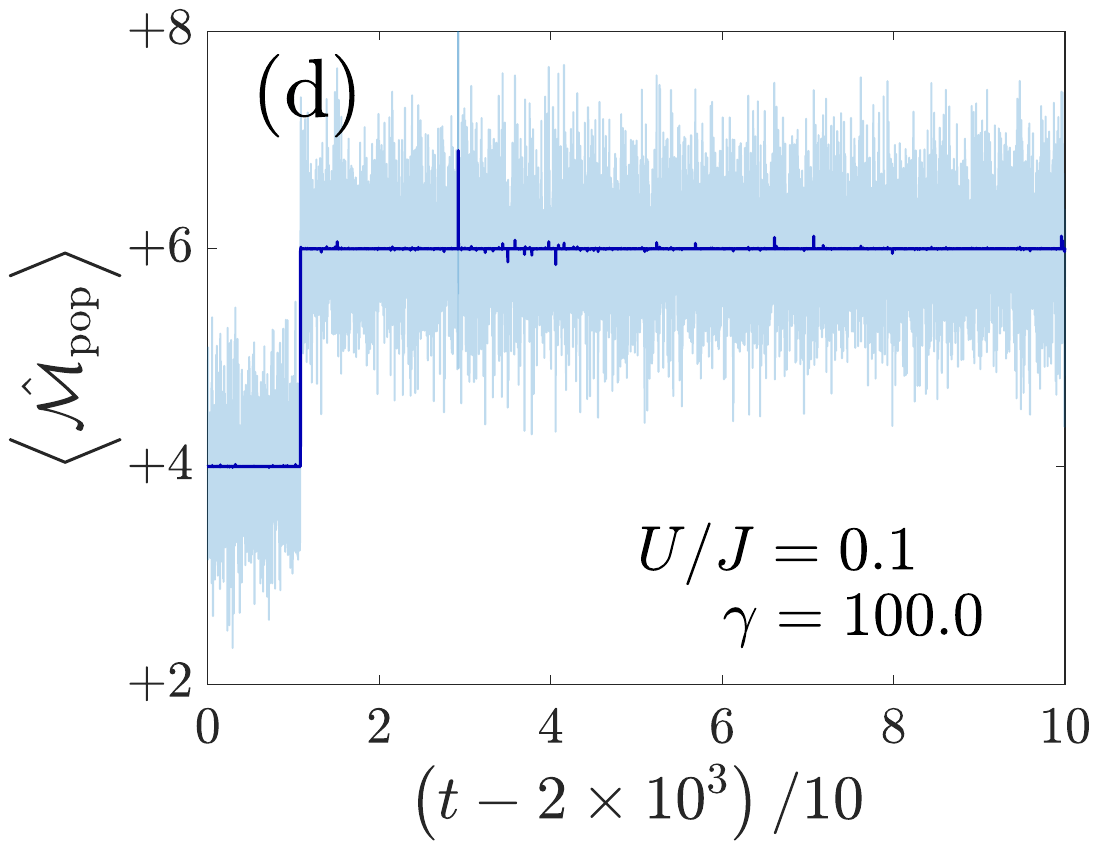}\hfill
\includegraphics[trim={0.5cm 7cm 0.5cm 7cm},clip, width=0.33\textwidth]{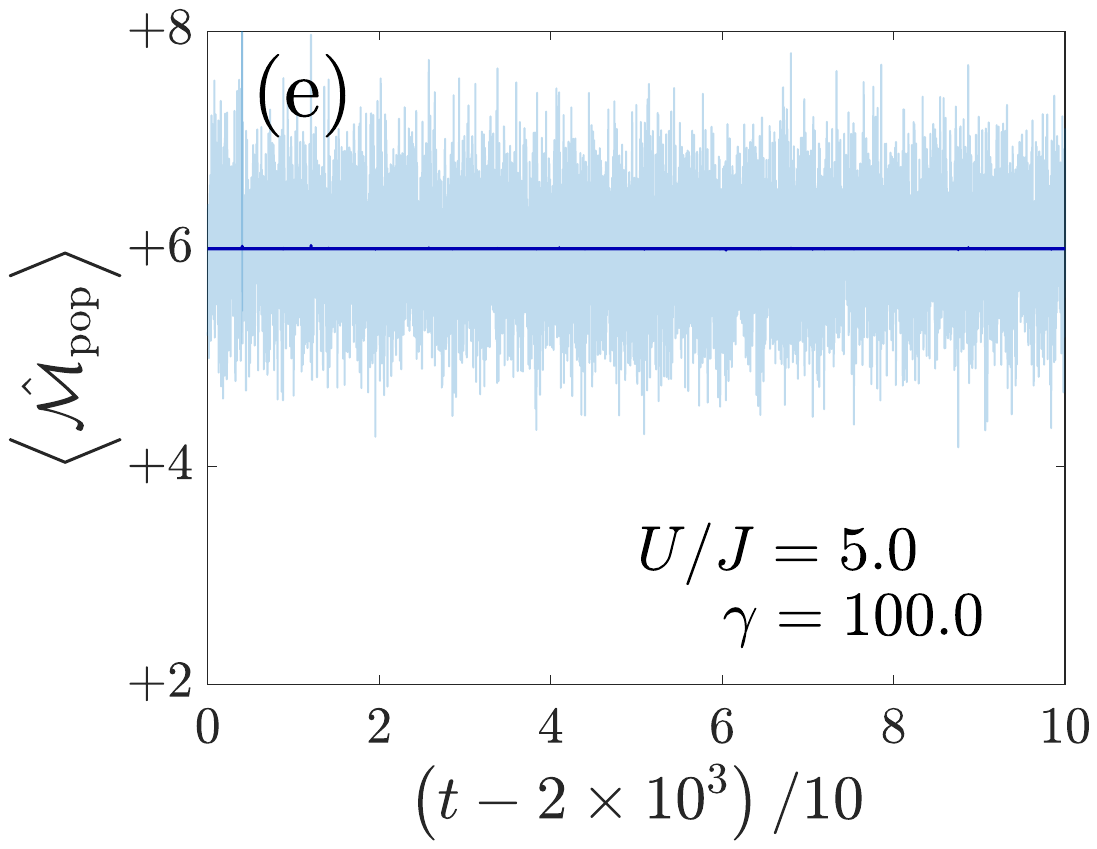}\hfill
\includegraphics[trim={0.5cm 7cm 0.5cm 7cm},clip, width=0.33\textwidth]{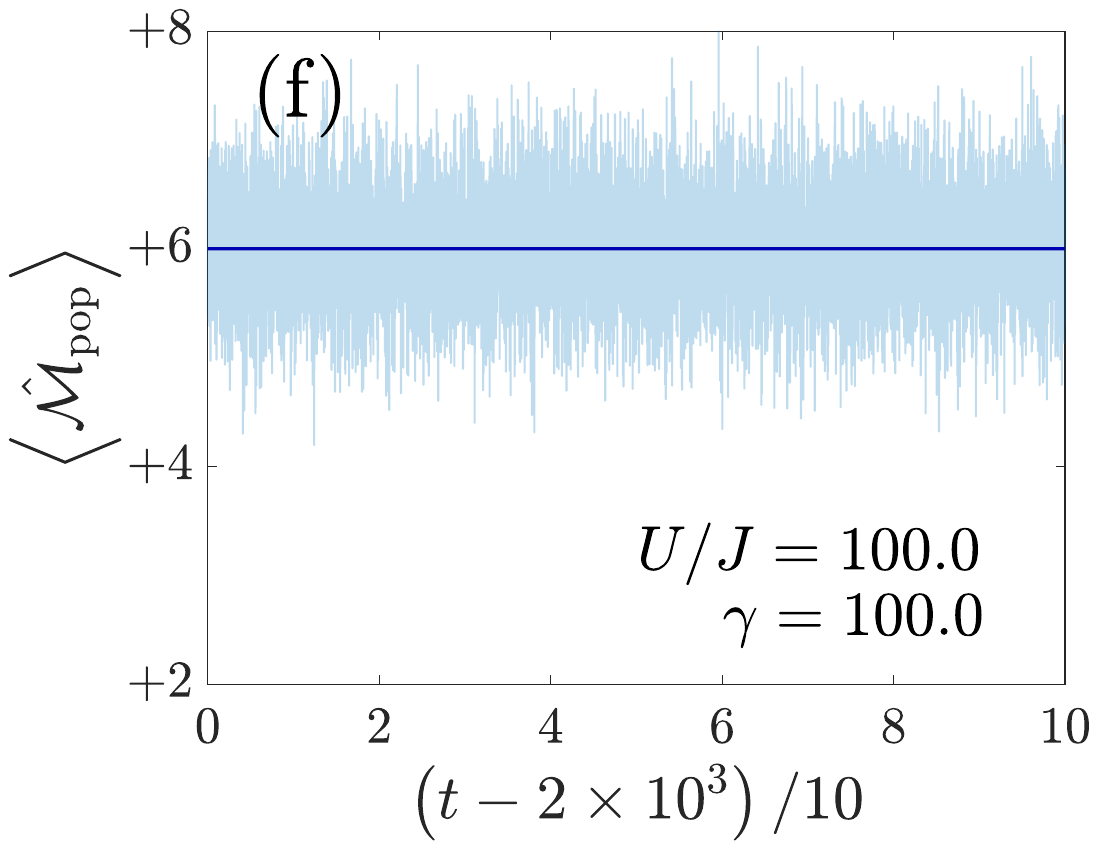}
\caption{We show the plots for the time dependent expectation value \eqref{ExptM0PlotDefn} of the measurement operator $\hat{\mathcal{M}}_\mathrm{pop}$ vs time and $I(t)/(2\gamma)$ vs time in dark and faded blue, respectively.  As expected, the superfluid dynamics here in \textbf{(a)} is similar to the Mott-insulator dynamics of Fig.\ \ref{FigCohDynamics}\textbf{(c)} and vice versa.  For the trajectories in the quantum Zeno regime \textbf{(d)}-\textbf{(f)}, we observe fewer quantum jumps here than in Figs.\ \ref{FigCohDynamics}\textbf{(d)}-\textbf{(f)}.}
\label{FigPopDynamics}
\end{figure}

\section{An Example: The Bose-Hubbard Model}
\label{Ex_BH}

To numerically demonstrate the stochastic Schr\"{o}dinger dynamics of Eq.\ \eqref{SSEPRA}, we start with the optical lattice potential
\begin{equation}
\label{OptLattBH}
V_l(x) = V_0\sin^{2}(kx),
\end{equation}
where the wave vector $k$ is related to the wavelength of the laser light by the relation $k = 2\pi/\lambda$.  Expanding the atomic field operators in terms of the Wannier functions for the optical lattice potential (\ref{OptLattBH}), one can show that $\hat{\mathcal{H}}_{\text{atm}}$ becomes the following Bose-Hubbard Hamiltonian \cite{BH}
\begin{equation}
\hat{\mathcal{H}}_{\text{BH}} = -J\sum_{\left\langle i,j \right\rangle} \hat{b}^{\dagger}_{i}\hat{b}_{j} + \sum_i \epsilon_i \hat{n}_i + \frac{U}{2}\sum_i  \hat{n}_i(\hat{n}_i - 1),
\label{BH_H}
\end{equation}
where $\hat{n}_i = \hat{b}^{\dagger}_{i}\hat{b}_{i}$ and the parameters of $\hat{\mathcal{H}}_{\text{BH}}$ are given by
\begin{subequations}
\begin{align}
J=&\int w_j(x)^*\left(-\frac{\hat{p}^2}{2m}+V_l(x)\right)w_{j+1}(x)\mathrm{d}x \\
U=& \;\;g_s\int |w_j(x)|^4 \mathrm{d}x, \\
\epsilon_i \approx& \;\;V_T(x_i),
\end{align}
\end{subequations}
with $V_l$ the lattice potential, $V_T$ a slowly varying external trapping potential, and $g_s$ the s-wave scattering length in one-dimension. 

Considering two different choices of the spatial mode function $f_a(x,\omega_L)$, we obtain the two Hermitian observables $\hat{\mathcal{M}}_0: \hat{\mathcal{M}}_\mathrm{coh}=m_\mathrm{coh}\sum_j\hat{b}_j^\dagger \hat{b}_{j+1}+h.c.$ and $\hat{\mathcal{M}}_0: \hat{\mathcal{M}}_\mathrm{pop}=m_\mathrm{pop}\sum_{j}\hat{b}_{2j}^\dagger \hat{b}_{2j}$, whose weak and continuous measurements was studied in Ref.\ \cite{APLBJJ}.  Here $m_\mathrm{coh}$ and $m_\mathrm{pop}$ are two constants calculated from the Wannier functions.  The Bose-Hubbard Hamiltonian \eqref{BH_H} has a superfluid to Mott-insulator transition at $U/J \approx 4.65$ in the thermodynamic limit \cite{BHED1}. The plot depicting the ground state expectation value of $\hat{\mathcal{M}}_\mathrm{coh}$ in Fig. \ref{FigCohOrderParameter} is consistent with this critical value.

In this section, we numerically study the stochastic Schr\"{o}dinger dynamics of Eq. \eqref{SSEPRA}. We show the examples of individual quantum trajectories due to the coherence measurement $\hat{\mathcal{M}}_\mathrm{coh}$ in Fig.\ \ref{FigCohDynamics} and due to the population measurement $\hat{\mathcal{M}}_\mathrm{pop}$ in Fig.\ \ref{FigPopDynamics}. In each of the panels of Figs.\ \ref{FigCohDynamics} and \ref{FigPopDynamics}, we plot the time dependent expectation value of the measurement operators 
\begin{equation}
\label{ExptM0PlotDefn}
\left\langle \hat{\mathcal{M}}_0 \right\rangle    =    \left\langle\Psi(t)\right|    \hat{\mathcal{M}}_0    \left|\Psi(t)\right\rangle  
\end{equation}
vs time for different values of the parameter $U/J$ and different measurement strength $\gamma$, where $\left|\Psi(t)\right\rangle$ is obtained numerically from the SSE \eqref{SSEPRA}. Since we are not interested in the transient effects, we have not shown the trajectory upto $t=2\times 10^3$.

From the plots shown in Figs.\ \ref{FigCohDynamics}(a)-(c) and \ref{FigPopDynamics}(a)-(c), we conclude that signatures of the superfluid to Mott-insulator transition can even be detected in the measurement signals (without the Wiener noise) for individual quantum trajectories. The phase transition is in agreement with the one detected in Ref.\ \cite{APLBJJ} with the help of power spectral densities. 

In Fig.\ \ref{FigCohDynamics}(a), we see that the coherence measurement signal remains almost constant in the superfluid phase with $U/J = 0.1$. On the other hand, in Fig.\ \ref{FigCohDynamics}(c), we observe chaotic oscillations in the Mott-insulating phase with $U/J = 100.0$. Interestingly, the coherence measurement appears to be able to distinguish the phases even in the strong measurement (quantum Zeno) regime at $\gamma = 10.0$. This is in stark contrast to the coherence measurement PSDs obtained for the quantum Zeno regime in Ref.\ \cite{APLBJJ}. However, one must point out that the nature of the stochastic jumps in Figs.\ \ref{FigCohDynamics}(e) and (f) are quite different from the chaotic fluctuations observed in Figs.\ \ref{FigCohDynamics}(b) and (c).   

In the weak measurement regime, one observes chaotic fluctuations in the population measurement signal in the superfluid phase in Fig.\ \ref{FigPopDynamics}(a) and almost constant signal in the Mott-insulator phase Fig.\ \ref{FigPopDynamics}(c). It is hard to distinguish the phases using the population signals in the quantum Zeno regime that are shown in \ref{FigPopDynamics}(d)-(f). Our study shows that typical quantum trajectories from different measurement signals contain a lot of information even in the quantum Zeno regime. 

In Ref.\ \cite{APLBJJ}, we explored ground state phase transitions in the Bose-Hubbard model via the power spectral density (PSD) of the homodyne measurement signal defined in Eq.\ \eqref{SSEPRA_Signal}. That analysis required a finite observation window of duration $T$, leading to a frequency resolution $\Delta f \sim 1/T$. While useful for identifying slow dynamical features, this approach masks short-timescale events --- especially under ensemble averaging. In the quantum Zeno regime, the PSDs for both coherence and population measurements appear similar, dominated by a peak at zero frequency. In this work, we instead analyze the measurement signal directly in the time domain, which preserves fast stochastic features. This reveals that, in the quantum Zeno regime, coherence measurements in the deep Mott-insulating phase exhibit frequent quantum jumps, whereas population measurements in the deep superfluid phase show such jumps only rarely.            

\section{Conclusion}
\label{SecConclude}

In this work, we have developed a detailed, microscopically grounded framework connecting a realizable homodyne detection scheme to the canonical form of Gaussian quantum continuous measurement for a strongly interacting many-body atomic system. Starting from a fully modeled atom-light Hamiltonian in a driven, lossy cavity, we systematically eliminated the excited atomic and cavity degrees of freedom to derive an effective Hamiltonian governing the interaction between the atomic observable and the input field. This led to an SSE expressed purely in terms of atomic operators. We explicitly delineated the approximations required to recover the canonical SSE form and emphasized that such a connection is nontrivial in the context of interacting systems. 

Numerical simulations using the Bose-Hubbard model further demonstrated the utility of this framework: while power spectral densities in the quantum Zeno regime show little variation across observables, time-resolved signals preserve key dynamical features such as the occurrence and frequency of quantum jumps.

Looking ahead, this formalism provides a robust foundation for probing measurement-induced phenomena in many-body systems. The clear mapping between a physical measurement scheme and the resulting conditional dynamics opens avenues for studying feedback protocols, dynamical phase transitions, and entanglement growth under continuous observation. Furthermore, our approach can be extended to explore non-Markovian regimes, where memory effects from the light-matter interface or structured environments play a significant role. Incorporating such effects while retaining a microscopically consistent treatment of measurement and detection will be crucial for advancing our understanding of open, interacting quantum systems at the interface of theory and experiment.

\ack
Aniket Patra acknowledges the financial support from the Institute for Basic Science (IBS) in the Republic of Korea through the project IBS-R024-D1. He is grateful for insightful discussions with Anne E. B. Nielsen.


\appendix

\section{Evaluation of Important Gaussian Integrals}
\label{Appn_Gaussian}

We show the details of how to compute the Gaussian integrals appearing in Eqs.\ \eqref{PkInit1} and (\ref{HomoCondDM2}) in this appendix. In both examples, we replace the summation over photon number $p$  by an integral with lower and upper limit of $0$ and $\infty$, respectively.

\subsection{Gaussian Integral Needed to Simplify the Probability $P_k$} 
\label{AppnGaussianPk}

Here we consider the summation that appears on the last line of Eq.\ \eqref{PkInit1}.  We replace the summation over $p$ by an integration and obtain
\begin{gather}
\label{SimplifyPk3Gaussian121}
\begin{aligned}
&\sum_{p = 0}^{\infty}\frac{1}{2\pi\left|\alpha_{m}^{\prime}\right|\left|\beta_{m}^{\prime}\right|}\exp{\left(-\frac{(p + k - \left|\alpha_{m}^{\prime}\right|^2)^{2}}{2\left|\alpha_{m}^{\prime}\right|^{2}} - \frac{(p-\left|\beta_{m}^{\prime}\right|^{2})^{2}}{2\left|\beta_{m}^{\prime}\right|^{2}}\right)} \\
&\approx \int_{0}^{\infty}\mathrm{d} p\frac{1}{2\pi\left|\alpha_{m}^{\prime}\right|\left|\beta_{m}^{\prime}\right|}\exp{\left(-\frac{(p + k -\left|\alpha_{m}^{\prime}\right|^2)^{2}}{2\left|\alpha_{m}^{\prime}\right|^2} - \frac{(p-\left|\beta_{m}^{\prime}\right|^2)^{2}}{2\left|\beta_{m}^{\prime}\right|^2}\right)} \\
&= \exp{\left[-\frac{\left(k + \left|\beta_{m}^{\prime}\right|^2 - \left|\alpha_{m}^{\prime}\right|^2\right)^{2}}{2\left(\left|\beta_{m}^{\prime}\right|^2 + \left|\alpha_{m}^{\prime}\right|^2\right)}\right]}
\times \int_{\frac{\left|\beta_{m}^{\prime}\right|^2\left(k - 2\left|\alpha_{m}^{\prime}\right|^2\right)}{\left|\beta_{m}^{\prime}\right|^2 + \left|\alpha_{m}^{\prime}\right|^2}}^{\infty}\mathrm{d} \widetilde{p}\frac{1}{2\pi\left|\alpha_{m}^{\prime}\right|\left|\beta_{m}^{\prime}\right|}\exp{\left[ -\left(\frac{\left|\beta_{m}^{\prime}\right|^2 + \left|\alpha_{m}^{\prime}\right|^2}{2\left|\beta_{m}^{\prime}\right|^2 \left|\alpha_{m}^{\prime}\right|^2}\right)\widetilde{p}^{2}\right]}, 
\end{aligned}
\end{gather}
where we have introduced a new shifted variable
\begin{equation}
\widetilde{p} = p + \frac{\left|\beta_{m}^{\prime}\right|^{2}\left(k - 2\left|\alpha_{m}^{\prime}\right|^2\right)}{\left|\beta_{m}^{\prime}\right|^{2} + \left|\alpha_{m}^{\prime}\right|^2}.
\label{pPrime}
\end{equation}  
Since $\left|\alpha_{m}^{\prime}\right|^2, \left|\beta_{m}^{\prime}\right|^2 \gg 1$, we have 
\begin{equation}
\label{GaussianLowerLimit1}
\frac{\left|\beta_{m}^{\prime}\right|^2\left(k - 2\left|\alpha_{m}^{\prime}\right|^2\right)}{\left|\beta_{m}^{\prime}\right|^2 + \left|\alpha_{m}^{\prime}\right|^2} \approx - \frac{    2\left|\alpha_{m}^{\prime}\right|^2\left|\beta_{m}^{\prime}\right|^2    }{    \left|\beta_{m}^{\prime}\right|^2 + \left|\alpha_{m}^{\prime}\right|^2    } \rightarrow -\infty.
\end{equation} 
Using the above, we extend the lower limit of the integration to $-\infty$ in  Eq.\ \eqref{SimplifyPk3Gaussian121} and finally obtain
\begin{gather}
\label{SimplifyPk3Gaussian122}
\begin{aligned}
&\sum_{p = 0}^{\infty}\frac{1}{2\pi\left|\alpha_{m}^{\prime}\right|\left|\beta_{m}^{\prime}\right|}\exp{\left(-\frac{(p + k - \left|\alpha_{m}^{\prime}\right|^2)^{2}}{2\left|\alpha_{m}^{\prime}\right|^{2}} - \frac{(p-\left|\beta_{m}^{\prime}\right|^{2})^{2}}{2\left|\beta_{m}^{\prime}\right|^{2}}\right)}  \\
&\approx \frac{1}{2\pi\left|\alpha_{m}^{\prime}\right|\left|\beta_{m}^{\prime}\right|}\exp{\left[-\frac{\left(k + \left|\beta_{m}^{\prime}\right|^2 - \left|\alpha_{m}^{\prime}\right|^2\right)^{2}}{2\left(\left|\beta_{m}^{\prime}\right|^2 + \left|\alpha_{m}^{\prime}\right|^2\right)}\right]}\int_{-\infty}^{+\infty}\mathrm{d} \widetilde{p}\exp{\left[ -\left(\frac{\left|\beta_{m}^{\prime}\right|^2 + \left|\alpha_{m}^{\prime}\right|^2}{2\left|\beta_{m}^{\prime}\right|^2 \left|\alpha_{m}^{\prime}\right|^2}\right)\widetilde{p}^{2}\right]} \\
&= \frac{1}{\sqrt{2\pi\left(\left|\beta_{m}^{\prime}\right|^2 + \left|\alpha_{m}^{\prime}\right|^2\right)}}\exp{\left[-\frac{\left(k + \left|\beta_{m}^{\prime}\right|^2 - \left|\alpha_{m}^{\prime}\right|^2\right)^{2}}{2\left(\left|\beta_{m}^{\prime}\right|^{2} + \left|\alpha_{m}^{\prime}\right|^2\right)}\right]}.
\end{aligned}
\end{gather}

\subsection{Gaussian Integral Needed to Simplify the Density Matrix $\rho(t + \mathrm{d} t)$} 
\label{AppnGaussianrhotdt}

We consider the summation that appears on the last line of Eq.\ \eqref{HomoCondDM2}. Replacing the summation over $p$ by an integration, we obtain   
\begin{gather}
\label{HomoCondDM21}
\begin{aligned}
&\sum_{p}    \exp \left[    -\frac{(p-\left|\beta_{m}^{\prime}\right|^2)^{2}}{4\left|\beta_{m}^{\prime}\right|^2}    -\frac{(p-\left|\beta_{n}^{\prime}\right|^2)^{2}}{4\left|\beta_{n}^{\prime}\right|^2}    -\frac{(p + k -\left|\alpha_{m}^{\prime}\right|^2)^{2}}{4\left|\alpha_{m}^{\prime}\right|^2}    -\frac{(p + k -\left|\alpha_{n}^{\prime}\right|^2)^{2}}{4\left|\alpha_{n}^{\prime}\right|^2}    - p\theta    \right] \\
&\approx \int_{0}^{\infty}\mathrm{d} p  \;\;\exp \left[    -\frac{(p-\left|\beta_{m}^{\prime}\right|^2)^{2}}{4\left|\beta_{m}^{\prime}\right|^2}    -\frac{(p-\left|\beta_{n}^{\prime}\right|^2)^{2}}{4\left|\beta_{n}^{\prime}\right|^2}    -\frac{(p + k -\left|\alpha_{m}^{\prime}\right|^2)^{2}}{4\left|\alpha_{m}^{\prime}\right|^2}    -\frac{(p + k -\left|\alpha_{n}^{\prime}\right|^2)^{2}}{4\left|\alpha_{n}^{\prime}\right|^2}    - p\theta    \right], 
\end{aligned}
\end{gather}
where $\theta$ will be computed in \ref{ValueOfITheta}. After completing the square in the argument of the exponential in Eq.\ \eqref{HomoCondDM21}, we note that
\begin{gather}
\label{CompleteSquare}
\begin{aligned}
&\frac{(p-\left|\beta_{m}^{\prime}\right|^2)^{2}}{4\left|\beta_{m}^{\prime}\right|^2}    +    \frac{(p-\left|\beta_{n}^{\prime}\right|^2)^{2}}{4\left|\beta_{n}^{\prime}\right|^2}    +    \frac{(p + k -\left|\alpha_{m}^{\prime}\right|^2)^{2}}{4\left|\alpha_{m}^{\prime}\right|^2}    +    \frac{(p + k -\left|\alpha_{n}^{\prime}\right|^2)^{2}}{4\left|\alpha_{n}^{\prime}\right|^2}    +    p\theta    \\
&= \frac{1}{4}\left[    \left( \left|\beta_{m}^{\prime}\right|^2 + \left|\beta_{n}^{\prime}\right|^2 + \left|\alpha_{m}^{\prime}\right|^2 + \left|\alpha_{n}^{\prime}\right|^2 \right)    -4k    + \frac{k^2}{\left|\alpha_{m}^{\prime}\right|^2} + \frac{k^2}{\left|\alpha_{n}^{\prime}\right|^2}   \right.    \\
&\left. - \frac{    \left|\beta_{m}^{\prime}\right|^2\left|\beta_{n}^{\prime}\right|^2    \left[  2\left|\alpha_{m}^{\prime}\right|^2 \left|\alpha_{n}^{\prime}\right|^2   \left(\theta - 2\right) + \left( \left|\alpha_{m}^{\prime}\right|^2 + \left|\alpha_{n}^{\prime}\right|^2 \right)k \right]^2    }{    \left|\alpha_{m}^{\prime}\right|^2 \left|\alpha_{n}^{\prime}\right|^2    \left[    \left|\alpha_{m}^{\prime}\right|^2 \left|\alpha_{n}^{\prime}\right|^2\left( \left|\beta_{m}^{\prime}\right|^2 + \left|\beta_{n}^{\prime}\right|^2 \right)    +    \left|\beta_{m}^{\prime}\right|^2  \left|\beta_{n}^{\prime}\right|^2\left(  \left|\alpha_{m}^{\prime}\right|^2 + \left|\alpha_{n}^{\prime}\right|^2 \right)   \right]    }     \right] \\
&+ \frac{1}{4}\left(  \frac{1}{\left|\beta_{m}^{\prime}\right|^2} + \frac{1}{\left|\beta_{n}^{\prime}\right|^2} + \frac{1}{\left|\alpha_{m}^{\prime}\right|^2} + \frac{1}{\left|\alpha_{n}^{\prime}\right|^2}  \right)     \left( \frac{    \left|\beta_{m}^{\prime}\right|^2  \left|\beta_{n}^{\prime}\right|^2    \left[  2\left|\alpha_{m}^{\prime}\right|^2 \left|\alpha_{n}^{\prime}\right|^2   \left(\theta - 2\right) + \left( \left|\alpha_{m}^{\prime}\right|^2 + \left|\alpha_{n}^{\prime}\right|^2 \right)k \right]    }{    \left|\alpha_{m}^{\prime}\right|^2 \left|\alpha_{n}^{\prime}\right|^2\left( \left|\beta_{m}^{\prime}\right|^2 + \left|\beta_{n}^{\prime}\right|^2 \right)    +    \left|\beta_{m}^{\prime}\right|^2  \left|\beta_{n}^{\prime}\right|^2\left(  \left|\alpha_{m}^{\prime}\right|^2 + \left|\alpha_{n}^{\prime}\right|^2 \right)    }    + p     \right)^{2}    \\
&\equiv Q_{1}(k) + Q_{20}\left( Q_{21} + p \right)^2,
\end{aligned}
\end{gather}
where in the last line we have defined new variables $Q_{1}, Q_{20},$ and $Q_{21}$. In particular, we define 
\begin{multline} 
\label{QkDefn}
Q_{1}(k) = \frac{1}{4}\left[    \left( \left|\beta_{m}^{\prime}\right|^2 + \left|\beta_{n}^{\prime}\right|^2 + \left|\alpha_{m}^{\prime}\right|^2 + \left|\alpha_{n}^{\prime}\right|^2 \right)    -4k    + \frac{k^2}{\left|\alpha_{m}^{\prime}\right|^2} + \frac{k^2}{\left|\alpha_{n}^{\prime}\right|^2}   \right.   \\\left. - \frac{    \left|\beta_{m}^{\prime}\right|^2\left|\beta_{n}^{\prime}\right|^2    \left[  2\left|\alpha_{m}^{\prime}\right|^2 \left|\alpha_{n}^{\prime}\right|^2   \left(\theta - 2\right) + \left( \left|\alpha_{m}^{\prime}\right|^2 + \left|\alpha_{n}^{\prime}\right|^2 \right)k \right]^2    }{    \left|\alpha_{m}^{\prime}\right|^2 \left|\alpha_{n}^{\prime}\right|^2    \left[    \left|\alpha_{m}^{\prime}\right|^2 \left|\alpha_{n}^{\prime}\right|^2\left( \left|\beta_{m}^{\prime}\right|^2 + \left|\beta_{n}^{\prime}\right|^2 \right)    +    \left|\beta_{m}^{\prime}\right|^2  \left|\beta_{n}^{\prime}\right|^2\left(  \left|\alpha_{m}^{\prime}\right|^2 + \left|\alpha_{n}^{\prime}\right|^2 \right)   \right]    }     \right]
\end{multline} 
to be a quadratic polynomial in $k$ that does not depend on $p$.  As a result, one can pull the factor $\exp\left[   -Q_{1}(k)   \right]$ out of the summation in Eq.\ \eqref{HomoCondDM21}. 

Using the variables introduced in Eq.\ \eqref{CompleteSquare}, we obtain from Eq.\ \eqref{HomoCondDM21}
\begin{gather}
\label{SimplifyHomoCondDMGaussian1}
\begin{aligned}
&\exp\left[   Q_{1}(k)   \right] \times   \sum_{p}    \exp \left[    -\frac{(p-\left|\beta_{m}^{\prime}\right|^2)^{2}}{4\left|\beta_{m}^{\prime}\right|^2}    -\frac{(p-\left|\beta_{n}^{\prime}\right|^2)^{2}}{4\left|\beta_{n}^{\prime}\right|^2}    -\frac{(p + k -\left|\alpha_{m}^{\prime}\right|^2)^{2}}{4\left|\alpha_{m}^{\prime}\right|^2}    -\frac{(p + k -\left|\alpha_{n}^{\prime}\right|^2)^{2}}{4\left|\alpha_{n}^{\prime}\right|^2}    - p\theta    \right]\\
&=    \int_{0}^{\infty} \mathrm{d} p     \;   \exp{\left[    -Q_{20}\left( Q_{21} + p \right)^2    \right]} \\
&= \int_{    Q_{21}^{\textrm{Re}}(k)    }^{\infty}\mathrm{d} \widetilde{p}_1    \;   \exp{\left[    -Q_{20}\left( Q_{21}^{\theta}(k) + \widetilde{p}_1 \right)^2    \right]},
\end{aligned}
\end{gather}
where we have introduced the variable change
\begin{equation}
\widetilde{p}_1    \equiv p    +    Q_{21}^{\textrm{Re}}(k)    =     p
+     \frac{    \left|\beta_{m}^{\prime}\right|^2  \left|\beta_{n}^{\prime}\right|^2    \left[  k\left( \left|\alpha_{m}^{\prime}\right|^2 + \left|\alpha_{n}^{\prime}\right|^2 \right) - 4\left|\alpha_{m}^{\prime}\right|^2 \left|\alpha_{n}^{\prime}\right|^2   \right]    }{    \left|\alpha_{m}^{\prime}\right|^2 \left|\alpha_{n}^{\prime}\right|^2\left( \left|\beta_{m}^{\prime}\right|^2 + \left|\beta_{n}^{\prime}\right|^2 \right)    +    \left|\beta_{m}^{\prime}\right|^2  \left|\beta_{n}^{\prime}\right|^2\left(  \left|\alpha_{m}^{\prime}\right|^2 + \left|\alpha_{n}^{\prime}\right|^2 \right)    }.
\label{pPrimePrime}
\end{equation}
The shift $Q_{21}^{\textrm{Re}}(k)$ becomes the new lower limit of the integral. Inside the integrand, we have introduced the new variable
\begin{equation}
Q_{21}^{\theta}(k)   \equiv   Q_{21}(k) - Q_{21}^{\textrm{Re}}(k)     =    \frac{  2 \theta \left|\alpha_{m}^{\prime}\right|^2 \left|\alpha_{n}^{\prime}\right|^2 \left|\beta_{m}^{\prime}\right|^2  \left|\beta_{n}^{\prime}\right|^2  }{  \left|\alpha_{m}^{\prime}\right|^2 \left|\alpha_{n}^{\prime}\right|^2\left( \left|\beta_{m}^{\prime}\right|^2 + \left|\beta_{n}^{\prime}\right|^2 \right)  +  \left|\beta_{m}^{\prime}\right|^2  \left|\beta_{n}^{\prime}\right|^2\left(  \left|\alpha_{m}^{\prime}\right|^2 + \left|\alpha_{n}^{\prime}\right|^2 \right)  }
\label{GaussianOffset}
\end{equation}
for convenience.  Since $\theta$ is complex,  $Q_{21}^{\theta}(k)$ is also a complex number. 

Similar to (\ref{GaussianLowerLimit1}), we can replace the lower limit of the integral $Q_{21}^{\textrm{Re}}(k)$ by $-\infty$ using the fact $\left|\alpha_{m}^{\prime}\right|^2,  \left|\alpha_{n}^{\prime}\right|^2, \left|\beta_{m}^{\prime}\right|^2, \left|\beta_{n}^{\prime}\right|^2 \gg 1$. This then obtains from Eq.\ \eqref{SimplifyHomoCondDMGaussian1} the following Gaussian integral offset by a $Q_{21}^{\theta}(k)$:
\begin{multline}
\label{SimplifyHomoCondDMGaussian2}
\int_{    Q_{21}^{\textrm{Re}}(k)    }^{\infty}\mathrm{d} \widetilde{p}_1 \;   \exp{\left[    -Q_{20}\left( Q_{21}^{\theta}(k) + \widetilde{p}_1 \right)^2    \right]} 
\approx \int_{    -\infty    }^{\infty}\mathrm{d} \widetilde{p}_1   \; \exp{\left[    -Q_{20}\left( Q_{21}^{\theta}(k) + \widetilde{p}_1 \right)^2    \right]}  \\
= \sqrt{   \frac{ \pi }{ Q_{20} }   }    =    \frac{    \sqrt{4\pi}    }{    \left(  \frac{1}{\left|\beta_{m}^{\prime}\right|^2} + \frac{1}{\left|\beta_{n}^{\prime}\right|^2} + \frac{1}{\left|\alpha_{m}^{\prime}\right|^2} + \frac{1}{\left|\alpha_{n}^{\prime}\right|^2}  \right)^{\frac{1}{2}}    },
\end{multline}
which produces the same result as a usual Gaussian integral. As a result, we finally have
\begin{multline}
\label{SimplifyHomoCondDMGaussian3}
\sum_{p}    \exp \left[    -\frac{(p-\left|\beta_{m}^{\prime}\right|^2)^{2}}{4\left|\beta_{m}^{\prime}\right|^2}    -\frac{(p-\left|\beta_{n}^{\prime}\right|^2)^{2}}{4\left|\beta_{n}^{\prime}\right|^2}    -\frac{(p + k -\left|\alpha_{m}^{\prime}\right|^2)^{2}}{4\left|\alpha_{m}^{\prime}\right|^2}    -\frac{(p + k -\left|\alpha_{n}^{\prime}\right|^2)^{2}}{4\left|\alpha_{n}^{\prime}\right|^2}    - p\theta    \right] \\
\approx    \frac{    \sqrt{4\pi}    }{    \left(  \frac{1}{\left|\beta_{m}^{\prime}\right|^2} + \frac{1}{\left|\beta_{n}^{\prime}\right|^2} + \frac{1}{\left|\alpha_{m}^{\prime}\right|^2} + \frac{1}{\left|\alpha_{n}^{\prime}\right|^2}  \right)^{\frac{1}{2}}    }\exp\left[   -Q_{1}(k)   \right], 
\end{multline}
where $Q_{1}(k) $ is defined in Eq.\ \eqref{QkDefn}.

\section{Approximate Value of $\theta$ Introduced in Eq.\ \eqref{ExpArg}}
\label{ValueOfITheta}

In this section, we compute the value of $\theta$ introduced in Eq.\ \eqref{ExpArg} while keeping terms upto $\mathcal{O}(\mathrm{d} t)$ and $\mathcal{O}\left(\alpha_0\beta/\left( \alpha^{2} + \beta^{2}\right)\right)$. Recall the definition
\begin{equation}
\label{1ExpArg}
\begin{split}
i\theta &= \mathrm{Arg} (\beta_{m}^{\prime}) + \mathrm{Arg} (\beta_{n}^{\prime *})  + \mathrm{Arg} (\alpha_{m}^{\prime}) + \mathrm{Arg} (\alpha_{n}^{\prime *}) \\
&= - \tan^{-1}\left( \frac{\beta + \alpha_0 m \mathrm{d} t}{\alpha} \right) + \tan^{-1}\left( \frac{\beta + \alpha_0 n \mathrm{d} t}{\alpha} \right) + \tan^{-1}\left( \frac{\beta - \alpha_0 m \mathrm{d} t}{\alpha} \right) - \tan^{-1}\left( \frac{\beta - \alpha_0 n \mathrm{d} t}{\alpha} \right),
\end{split}
\end{equation}
where $-\pi/2  \leqslant  \mathrm{Arg} (\beta_{m}^{\prime}), \mathrm{Arg} (\beta_{n}^{\prime *}), \mathrm{Arg} (\alpha_{m}^{\prime}), \mathrm{Arg} (\alpha_{n}^{\prime *}) <+\pi/2$. Using the formula
\begin{equation}
\tan^{-1}(x) - \tan^{-1}(y) = \tan^{-1}\left( \frac{x - y}{1 + xy} \right)
\end{equation}
and neglecting terms  $\mathcal{O}(\mathrm{d} t^2)$ in the denominator, we obtain
\begin{equation}
\label{2ExpArg}
- \tan^{-1}\left( \frac{\beta + \alpha_0 m \mathrm{d} t}{\alpha} \right) + \tan^{-1}\left( \frac{\beta + \alpha_0 n \mathrm{d} t}{\alpha} \right)
= \tan^{-1}\left(  \frac{\alpha\alpha_0(n-m)\mathrm{d} t}{  \alpha^{2} + \beta^{2} +  \alpha_{0}\beta (m+n)\mathrm{d} t  }  \right).
\end{equation}
Simplifying the other two terms in Eq.\ \eqref{1ExpArg} similarly, we write
\begin{equation}
\label{3ExpArg}
\begin{split}
i\theta &= \tan^{-1}\left(  \frac{\alpha\alpha_0(n-m)\mathrm{d} t}{  \alpha^{2} + \beta^{2} +  \alpha_{0}\beta (m+n)\mathrm{d} t  }  \right) + \tan^{-1}\left(  \frac{\alpha\alpha_0(n-m)\mathrm{d} t}{  \alpha^{2} + \beta^{2} -  \alpha_{0}\beta (m+n)\mathrm{d} t  }  \right) \\
&\approx \tan^{-1}\left[ \frac{\alpha\alpha_0(n-m)\mathrm{d} t}{ \alpha^{2} + \beta^{2}  }  \left(  1 - \frac{\alpha_{0}\beta (m+n)\mathrm{d} t}{\alpha^{2} + \beta^{2}}  \right)  \right]  +  \tan^{-1}\left[ \frac{\alpha\alpha_0(n-m)\mathrm{d} t}{ \alpha^{2} + \beta^{2}  }  \left(  1 + \frac{\alpha_{0}\beta (m+n)\mathrm{d} t}{\alpha^{2} + \beta^{2}}  \right)  \right] \\
&\approx  2\tan^{-1}\left( \frac{\alpha\alpha_0(n-m)\mathrm{d} t}{ \alpha^{2} + \beta^{2}  }   \right)  \\
&\approx 2\left( \frac{\alpha\alpha_0(n-m)\mathrm{d} t}{ \alpha^{2} + \beta^{2}  }  \right).
\end{split}
\end{equation}
In Eq.\ \eqref{3ExpArg} we have used the following Taylor series expansions for $-1\leqslant x \leqslant +1$:
\begin{equation}
\tan^{-1}(x) = \sum_{n = 0}^{\infty}(-1)^n\frac{x^{2n + 1}}{2n + 1}, \qquad \frac{1}{1 - x} = \sum_{n = 0}^{\infty}x^n,
\end{equation}
and kept terms upto $\mathcal{O}(\mathrm{d} t)$.

\section{Simplification of the Density Matrix $\rho(t + \mathrm{d} t)$} 
\label{AppnDM}

In this section, we simplify $\rho(t + \mathrm{d} t)$ and finally write it as a rank-1 projector.  Using Eq.\ \eqref{SimplifyHomoCondDMGaussian3} in Eq.\ \eqref{HomoCondDM2}, we derive 
\begin{gather}
\label{HomoCondDM3}
\begin{aligned}
\rho(t+\mathrm{d} t) &= \left|\Psi(t+\mathrm{d} t)\right\rangle\left\langle\Psi(t+\mathrm{d} t)\right| \\
&\approx \sum_{m,n}    \frac{    \sqrt{4\pi}    C_{m}C_{n}^{*}e^{-i\alpha_0\alpha (m-n) \mathrm{d} t}             \left[        e^{        -i    \hat{\widetilde{\mathcal{H}}}_{\text{atm}}          \mathrm{d} t        }        \left|\psi_{m}\right\rangle \left\langle\psi_{n}\right|        e^{        +i        \hat{\widetilde{\mathcal{H}}}_{\text{atm}}          \mathrm{d} t        }       \right]               }{    2\pi P_{k} \left[\left|\beta_{m}^{\prime}\right|^2\left|\beta_{n}^{\prime}\right|^2 \left|\alpha_{m}^{\prime}\right|^2\left|\alpha_{n}^{\prime}\right|^2\right]^{\frac{1}{4}}    \left(  \frac{1}{\left|\beta_{m}^{\prime}\right|^2} + \frac{1}{\left|\beta_{n}^{\prime}\right|^2} + \frac{1}{\left|\alpha_{m}^{\prime}\right|^2} + \frac{1}{\left|\alpha_{n}^{\prime}\right|^2}  \right)^{\frac{1}{2}}    }    \exp \left[    -Q_{1}(k)      \right]    \\
&\approx              \sum_{m,n}    \frac{    C_{m}C_{n}^{*}       e^{      -i\alpha_0\alpha (m-n) \mathrm{d} t       }         \left[        e^{        -i    \hat{\widetilde{\mathcal{H}}}_{\text{atm}}          \mathrm{d} t        }        \left|\psi_{m}\right\rangle \left\langle\psi_{n}\right|        e^{        +i        \hat{\widetilde{\mathcal{H}}}_{\text{atm}}          \mathrm{d} t        }       \right]         }{      P_k    \left[    2\pi\left( \alpha^{2} + \beta^{2}\right)    \right]^{\frac{1}{2}    }    }    \exp \left[    -Q_{1}(k)      \right],
\end{aligned}
\end{gather}
where we have used 
\begin{equation}
\label{HomoCondDM31}
\left|\beta_{m}^{\prime}\right|^2\left|\beta_{n}^{\prime}\right|^2 \left|\alpha_{m}^{\prime}\right|^2\left|\alpha_{n}^{\prime}\right|^2 \approx \frac{1}{16}\left( \alpha^{2} + \beta^{2}\right)^4,      \qquad      \left(  \frac{1}{\left|\beta_{m}^{\prime}\right|^2} + \frac{1}{\left|\beta_{n}^{\prime}\right|^2} + \frac{1}{\left|\alpha_{m}^{\prime}\right|^2} + \frac{1}{\left|\alpha_{n}^{\prime}\right|^2}  \right) \approx \frac{8}{\alpha^{2} + \beta^{2}},
\end{equation}
after neglecting $\mathcal{O}(\mathrm{d} t^2)$ terms.  

We now simplify $Q_{1}(k)$ from Eq.\ \eqref{QkDefn} by neglecting $\mathcal{O}(\mathrm{d} t^2)$ terms as 
\begin{gather}
\label{QkDefnSimple}
\begin{aligned}
Q_{1}(k) &\approx \frac{    k^2   }{   \alpha^{2} + \beta^{2}   }\left( \frac{1}{2} - \frac{4\alpha_0\beta (m + n) \mathrm{d} t}{\alpha^{2} + \beta^{2}}  \right) + k\frac{    \left\lbrace  i\alpha_0\alpha (n - m) \mathrm{d} t  +  \alpha_0\beta (m + n) \mathrm{d} t \right\rbrace    }{   \left(  \alpha^{2} + \beta^{2}  \right)   }    -i\alpha_0\alpha (n - m) \mathrm{d} t \\
& \overset{*1}{\approx} \frac{    k^2     +     2k\left\lbrace  i\alpha_0\alpha (n - m) \mathrm{d} t  +  \alpha_0\beta (m + n) \mathrm{d} t \right\rbrace    }{2\left(  \alpha^{2} + \beta^{2}  \right)}    -i\alpha_0\alpha (n - m) \mathrm{d} t \\
& \overset{*2}{\approx} \frac{    k^2     +     2k  \alpha_0\beta (m + n) \mathrm{d} t    }{2\left(  \alpha^{2} + \beta^{2}  \right)}    -i\alpha_0\alpha (n - m) \mathrm{d} t. 
\end{aligned}
\end{gather}
Since $\beta^2 = \mathcal{O}(1/\mathrm{d} t) \gg 1$, in the second line ($*1$) of Eq.\ \eqref{QkDefnSimple} we have neglected the term that is proportional to $\alpha_0\beta/\left( \alpha^{2} + \beta^{2}\right)^2$ in the coefficient of $k^2$.  Moreover, since the probe light is quite weak compared to the local oscillator, we also neglect $\alpha_0\alpha/\left( \alpha^{2} + \beta^{2}\right)$ compared to $\alpha_0\beta/\left( \alpha^{2} + \beta^{2}\right)$ in the final line ($*2$).

First neglecting the constant factor $P_k \sqrt{\left[2\pi\left( \alpha^{2} + \beta^{2}\right)\right]}$ and then using Eq.\ \eqref{QkDefnSimple} into (\ref{HomoCondDM3}), we obtain 
\begin{gather}
\label{HomoCondDM4}
\begin{aligned}
\rho(t+\mathrm{d} t) &= \left|\Psi(t+\mathrm{d} t)\right\rangle\left\langle\Psi(t+\mathrm{d} t)\right| \\
&\propto     \sum_{m,n}    C_{m}C_{n}^{*}e^{-i\alpha_0\alpha (m-n) \mathrm{d} t}         \left[        e^{        -i    \hat{\widetilde{\mathcal{H}}}_{\text{atm}}          \mathrm{d} t        }        \left|\psi_{m}\right\rangle \left\langle\psi_{n}\right|        e^{        +i        \hat{\widetilde{\mathcal{H}}}_{\text{atm}}          \mathrm{d} t        }       \right]          \exp \left[    -Q_{1}(k)      \right] \\
&\approx \sum_{m,n}    C_{m}C_{n}^{*}e^{-i\alpha_0\alpha (m-n) \mathrm{d} t}  e^{    \left[    -      \frac{    k^2     +     2k \alpha_0\beta (m + n) \mathrm{d} t  }{2\left(  \alpha^{2} + \beta^{2}  \right)}    + i\alpha_0\alpha (n - m) \mathrm{d} t   \right]  }          \left[        e^{        -i    \hat{\widetilde{\mathcal{H}}}_{\text{atm}}          \mathrm{d} t        }        \left|\psi_{m}\right\rangle \left\langle\psi_{n}\right|        e^{        +i        \hat{\widetilde{\mathcal{H}}}_{\text{atm}}          \mathrm{d} t        }       \right]         \\
&\approx    \sum_{m,n}    \left[    C_{m}C_{n}^{*} e^{        -i    \hat{\mathcal{H}}^\text{eff}_{\text{atm}  }         \mathrm{d} t        }        \left|\psi_{m}\right\rangle \left\langle\psi_{n}\right|        e^{        +i       \hat{\mathcal{H}}^\text{eff}_{\text{atm}  }          \mathrm{d} t        }        \right]    \exp\left[    -    \frac{\left\lbrace    2k^2 + 4k\alpha_0\beta (m + n) \mathrm{d} t    \right\rbrace}{  4(\alpha^{2} + \beta^{2})  }        \right] \\
&\approx \sum_{m,n}    C_{m}C_{n}^{*}     e^{  -i\hat{\mathcal{H}}^\text{eff}_{\text{atm} }\mathrm{d} t}  e^{-\frac{\left[k + 2\alpha_0\beta m \mathrm{d} t\right]^{2}}{4\left[ \alpha^{2} + \beta^{2}\right]}}  \left|\psi_{m}\right\rangle \left\langle\psi_{n}\right|  e^{-\frac{\left[k + 2\alpha_0\beta n \mathrm{d} t\right]^{2}}{4\left[ \alpha^{2} + \beta^{2}\right]}}  e^{  +i\hat{\mathcal{H}}^\text{eff}_{\text{atm}  }\mathrm{d} t}   \\
&=    e^{-i\hat{\mathcal{H}}^\text{eff}_{\text{atm}}\mathrm{d} t}  e^{-\frac{\left[\hat{k} + 2\alpha_0\beta \hat{\mathcal{M}} \mathrm{d} t\right]^{2}}{4\left[ \alpha^{2} + \beta^{2}\right]}}  \left[\sum_{m,n}C_{m}C_{n}^{*}\left|\psi_{m}\right\rangle \left\langle\psi_{n}\right|\right]  e^{-\frac{\left[\hat{k} + 2\alpha_0\beta \hat{\mathcal{M}} \mathrm{d} t\right]^{2}}{4\left[ \alpha^{2} + \beta^{2}\right]}}  e^{+i\hat{\mathcal{H}}^\text{eff}_{\text{atm}}\mathrm{d} t}    \\
&= e^{  -i    \hat{\mathcal{H}}^\text{eff}_{\text{atm}  }    \mathrm{d} t}  e^{-\frac{\left[\hat{k} + 2\alpha_0\beta \hat{\mathcal{M}} \mathrm{d} t\right]^{2}}{4\left[ \alpha^{2} + \beta^{2}\right]}}  \left|\Psi(t)\right\rangle\left\langle\Psi(t)\right|  e^{-\frac{\left[\hat{k} + 2\alpha_0\beta \hat{\mathcal{M}} \mathrm{d} t\right]^{2}}{4\left[ \alpha^{2} + \beta^{2}\right]}}  e^{  +i\hat{\mathcal{H}}^\text{eff}_{\text{atm}  }\mathrm{d} t},
\end{aligned}
\end{gather}
where 
\begin{equation}
\hat{\mathcal{H}}_\text{atm}^\text{eff}     =     \hat{\widetilde{\mathcal{H}}}_{\text{atm}} + 2 \alpha_0\alpha \hat{\mathcal{M}}.
\label{UtmostFinalEffH}
\end{equation}


\section*{References}


\end{document}